\documentclass[aps, reprint,amsmath,amssymbols,superscriptaddress,nofootinbib]{revtex4-1}

\usepackage{graphicx}
\usepackage{bm}
\usepackage{xcolor}
\usepackage{soul}
\usepackage{multirow,hhline}
\usepackage{lipsum}

\definecolor{darkBlue}{rgb}{0,0,0.6}
\definecolor{darkRed}{rgb}{0.5,0,0}
\definecolor{darkGreen}{rgb}{0,0.5,0}

\begin{document}


\title{Vortex dynamics and losses due to pinning:
Dissipation from trapped magnetic flux in resonant
superconducting radio-frequency cavities}
\author{Danilo B. Liarte}
\affiliation{Laboratory of Atomic and Solid State Physics, Cornell University, Ithaca, NY, USA}
\author{Daniel Hall}
\affiliation{Cornell Laboratory for Accelerator-Based Sciences and Education,  Cornell University,
Ithaca, NY, USA}
\author{Peter N. Koufalis}
\affiliation{Cornell Laboratory for Accelerator-Based Sciences and Education,  Cornell University,
Ithaca, NY, USA}
\author{Akira Miyazaki}
\affiliation{CERN, Geneva, Switzerland}
\affiliation{University of Manchester, Manchester, UK}
\author{Alen Senanian}
\affiliation{Laboratory of Atomic and Solid State Physics, Cornell University, Ithaca, NY, USA}
\author{Matthias Liepe}
\affiliation{Cornell Laboratory for Accelerator-Based Sciences and Education,  Cornell University,
Ithaca, NY, USA}
\author{James P. Sethna}
\affiliation{Laboratory of Atomic and Solid State Physics, Cornell University, Ithaca, NY, USA}




\date{\today}

\begin{abstract}
We use a model of vortex dynamics and collective weak pinning theory to study
the residual dissipation due to trapped magnetic flux in a dirty superconductor.
Using simple estimates, approximate analytical calculations, and numerical
simulations, we make predictions and comparisons with experiments performed
in CERN and Cornell on resonant superconducting radio-frequency NbCu,
doped-Nb and Nb$_3$Sn cavities.
We invoke hysteretic losses originating in a rugged pinning potential landscape
to explain the linear behavior of the sensitivity of the residual resistance to
trapped magnetic flux as a function of the amplitude of the radio-frequency field.
Our calculations also predict and describe the crossover from
hysteretic-dominated to viscous-dominated regimes of dissipation.
We propose simple formulas describing power losses and crossover behavior,
which can be used to guide the tuning of material parameters to optimize cavity
performance.
\end{abstract}

\pacs{}

\maketitle


\section{Introduction}
\label{sec:introduction}
Vortex matter is the ``smoking gun'' of type II superconductors%
~\cite{blatter94,brandt95,huebener01,blatter03}, typically appearing in
the form of a lattice of quantized magnetic flux lines in equilibrium
superconductor states at intermediate ranges of applied magnetic fields
and low temperatures.
Compared to clean Meissner states, the vortex state is not a \emph{good}
superconductor state; transverse transport currents ($\bm{j} \perp \bm{H}$,
with $\bm{H}$ representing the vortex magnetic field) acting on the vortex
flux line via Lorentz forces can dissipate power.
To restore dissipation-free current flows and control the dissipation of
high-temperature superconductors, it has become common practice to employ
impurity doping to pin the vortices and restrain their motion.
A dirty superconductor is often a \emph{good} superconductor.
Incidentally, high-power resonant superconducting radio-frequency (SRF)
cavities for particle accelerators operate in the metastable Meissner state%
~\cite{padamsee08,padamsee09,posen15}, i.e. at magnetic fields above the
lower critical field and below the superheating field~\cite{liarte17}, which might
mislead one to conclude that vortex motion have negligible, if any impact
on cavity power dissipation.
Here pinning by impurities plays a double role.
On the one hand, defects can trap vortex flux lines (originating in the earth
magnetic fields, thermo-electric currents, etc) that should have been expelled
from the superconductor during the cavity cool-down.
On the other hand, pinning can restrain the motion of the trapped vortices and
restore the desired dissipation-free current flow property of the Meissner state.
In typical SRF applications, oscillating magnetic fields parallel to the
superconductor interface can move isolated flux lines near the surface, and
produce non-negligible contributions to the cavity surface resistance.

In this paper, we use a model of vortex dynamics and collective weak pinning
theory~\cite{blatter03} to study the dissipation of an isolated superconducting
vortex line in a Gaussian random disordered potential  (due to weak pinning
on defects), subject to a time-dependent forcing near the surface (due to the
alternating magnetic fields $B_\text{rf}$ parallel to the inner surface of the
SRF cavity).
We will compare our results to three experimental measurements, for doped
Nb, Nb$_3$Sn, and NbCu cavities.%
  \footnote{For those not in the accelerator community, NbCu cavities are niobium 
  films a few microns thick sputtered onto copper substrates -- not a compound.}

Superconductors subject to oscillating fields dissipate power on their surface
due to thermal excitation of quasi-particles, even if there is no vortex matter.
We write the surface resistance of a superconductor as~\cite{padamsee08}
\begin{equation}
    R_S = \frac{2}{{H_{rf}}^2} \, P,
    \label{eq:surfaceR}
\end{equation}
where $P$ is the power per unit area dissipated in the superconductor wall
and $H_{rf}$ is the amplitude of the rf applied magnetic
field%
\footnote{Henceforth, we restrict our attention to ac rf fields.}.
The surface resistance decomposes into temperature-dependent and
temperature-independent parts, $R_S = R_{BCS}+ R_0$, with the former
and latter named BCS and \emph{residual} resistance, respectively.
The BCS part is usually described by BCS theory%
\footnote{
	The decomposition into temperature-dependent and independent parts
	is phenomenological, and the linear response of BCS theory
	(Mattis-Bardeen) does not necessarily describe some experimental
	results (see~\cite{miyazaki18}).
}.

The residual part is caused by several factors.
Here we focus our attention to the case where $R_0$ is caused primarily by
trapped magnetic flux.
Indeed, recent measurements in current cavity designs show that the
temperature independent residual resistance $R_0$ can be a large fraction
of the total dissipation (from about 20\% for Nb to 50\% for Nb$_3$Sn)
at operating temperatures~\cite{posen14b,gonnella16b} and
that it is roughly linear in the density of trapped flux~\cite{gonnella16b}.
The fact that $R_0$ is negligible at small trapped flux strongly suggests
that it is due to vortices; the linearity suggests that the vortices are not
interacting strongly with one another -- motivating our study of the dissipation
due to a single flux line.
Measurements of trapped-flux residual resistance are routinely employed
by the SRF community to quantify power losses and cavity quality factors.
Typical experiments show a characteristic bell shape dependence of $R_0$
as a function of the electronic mean free path~\cite{gonnella16,checchin17},
though early Nb films display a still intriguing ``U''-shaped dependence%
~\cite{benvenuti99}.

Previous theoretical calculations of dissipation~\cite{gurevich13,checchin17}
have ignored the effects of collective weak pinning on vortex motion, and have
derived a value for the residual resistance $R_0$ that is independent of
the amplitude $B_\text{rf}$ of the cavity rf field.
The recent cavities show a residual resistance that is roughly linear in the
rf field $B_\text{rf}$ (and hence a dissipation that is cubic in the rf field)%
~\cite{hall17,miyazaki18,miyazaki_TTCFermi_17,checchin18}.
Also, our calculations show that the total dissipation, ignoring pinning,
predicts not only a constant $R_0$, but one that is much higher than the 
measured dissipation at low fields%
  \footnote{The previous theories fix the vortex position at a certain depth
  in the material (corresponding to a single, inescapable pinning point), which 
  can be used to reduce the dissipation, but cannot introduce a dependence 
  of the dissipation on the strength of the rf field.}.
Since the energy dissipated by a moving vortex%
  \footnote{This ignores the contribution of quasiparticle excitations inside
  the vortex core, which contribute a small constant term to $R_0$.}
is given by the Lorentz force times the distance moved at the surface, some
kind of pinning must be included to restrict the amplitude of motion.
This motivates our consideration of collective weak pinning.
We shall find that collective weak pinning does indeed predict a linear
dependence of $R_0$ on $B_\text{rf}$.
Our estimates suggest that weak pinning due to point impurities (dopants)
is a factor of 6-20 too small to explain the low losses observed, and will
discuss the possible role of extended defects (dislocations, grain boundaries)
and  other possible reasons for the remaining discrepancy.

\begin{figure}[!ht]
\begin{minipage}[t]{0.4\linewidth}
\vspace{0.2cm}
\centering (a) \par\smallskip
\includegraphics[width=\linewidth]{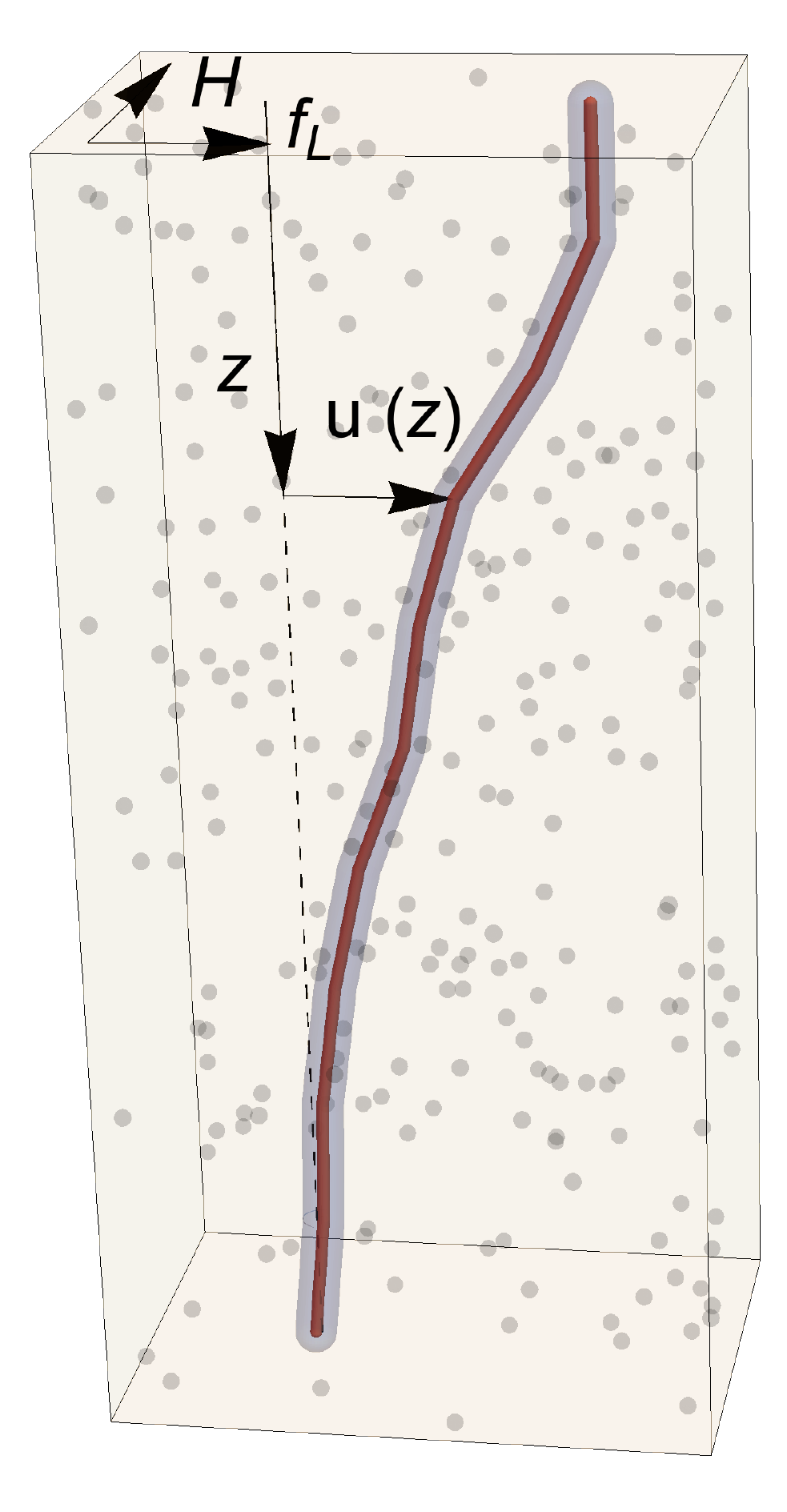}
\end{minipage}
\begin{minipage}[t]{0.57\linewidth}
\centering (b) \par\smallskip
\vspace{-0.5cm}
\includegraphics[width=\linewidth]{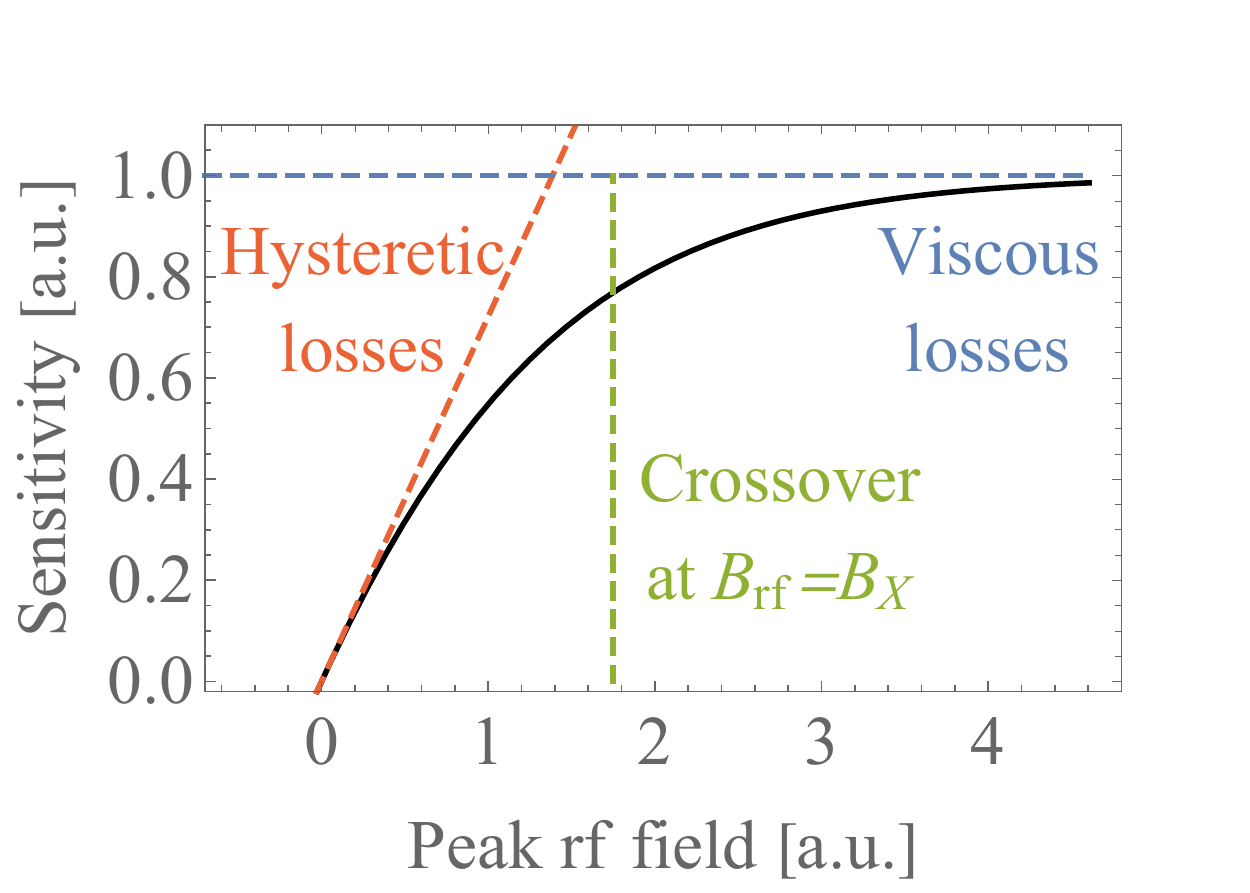}
\centering (c) \par\smallskip
\includegraphics[width=\linewidth]{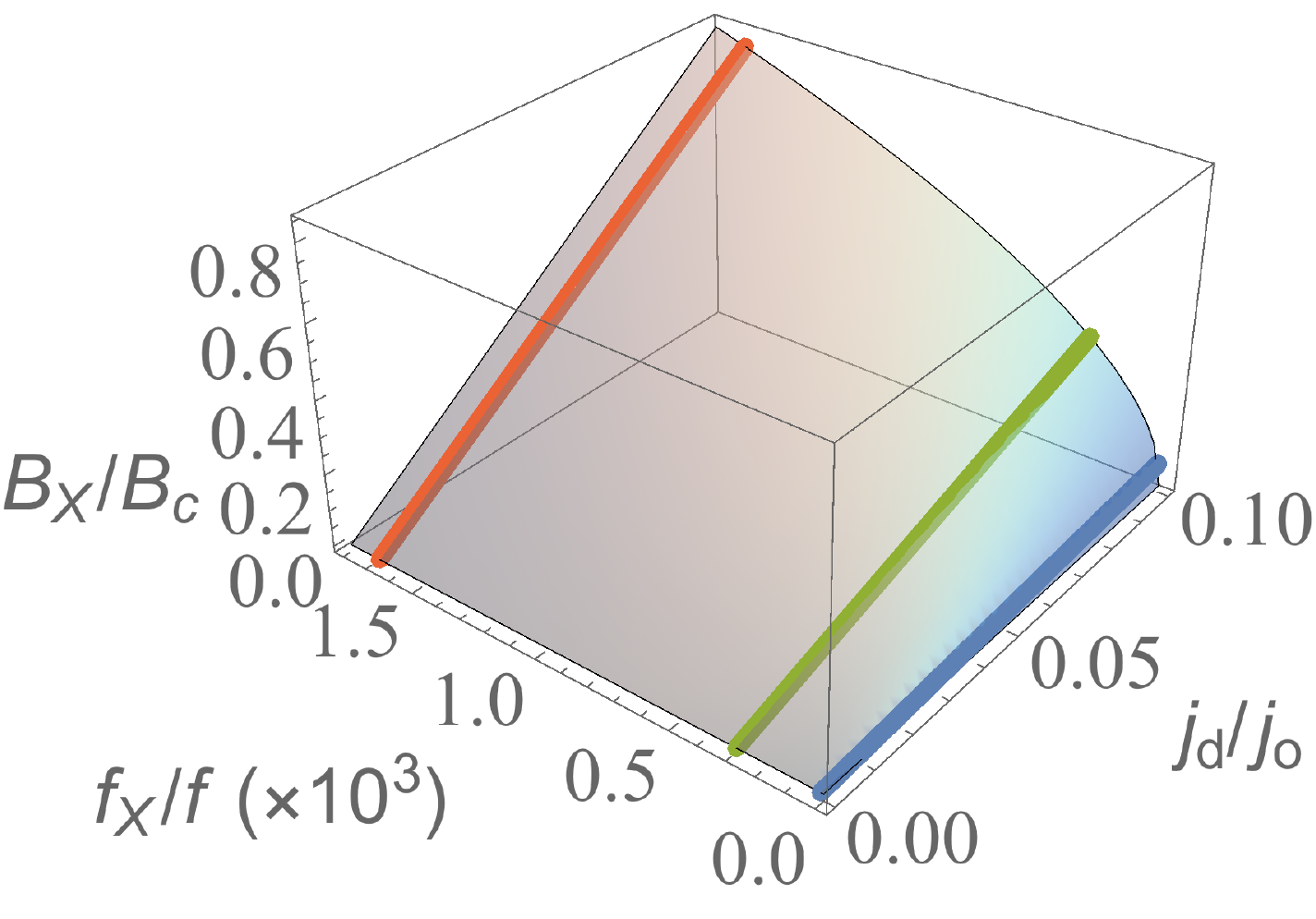}
\end{minipage}
\caption{%
(a) Illustration of an elastic vortex line subject to a Lorentz force and
random pinning forces.
(b) Sketch of our theoretical predictions for the sensitivity of the residual
resistance to trapped flux as a function of the amplitude of the rf magnetic
field.
(c) Dimensionless crossover field as a function of the depinning current
and inverse frequency.
\label{fig:summary}}
\end{figure}

It is surprising that the dynamical behavior of an individual vortex is less
well-known and understood than that of many interacting vortices%
\footnote{Although individual vortex lines are easier to study, there are
just a few situations, such as trapped flux dissipation in SRF cavities,
in which they play important roles.}%
~\cite{auslaender09}.
To study dynamics, we consider an idealized model where the vortex line
is an elastic one-dimensional string whose conformation is fully described
by a displacement field $u=u(z)$ from a reference configuration, where $z$
is the Cartesian coordinate associated with the distance from the
superconductor surface, and we assume $u(z)=0 \, \forall z$ in the
reference configuration (see Fig~\ref{fig:summary}a).
The displacement field satisfies the equation of motion,
\begin{equation}
f_V + f_E + f_L + f_P=0,
\label{eq:dynamics1}
\end{equation}
where $f_I$ denotes a force per length, and the subscripts $V,$ $E,$ $L,$
and $P$ are associated with viscous, elastic, Lorentz, and pinning forces,
respectively%
\footnote{In this paper, we neglect inertial and Magnus forces, which have
sub-dominant contributions.}.
Gurevich and Ciovati studied the ac dynamics of individual vortex lines
\emph{strongly}, irreversibly pinned at fixed distances from the interface,
and made contact with thermal measurements of hot spots in Nb
cavities~\cite{gurevich13}.
They assume $f_P=0$, and implement strong pinning by fixing one end of
the vortex line so that  $u(\ell_P)=0$ for a pinning center at $z=\ell_P.$
More recently, Checchin et al. extended the Gittleman-Rosenblum
model~\cite{gittleman66} to study \emph{weakly}, but also irreversibly pinned
vortices using the harmonic approximation for the pinning potential and
neglecting the vortex line tension $f_E$~\cite{checchin17}.
Working with cuprates (YBCO), Auslaender et al. used collective weak
pinning theory to study low-frequency dynamic properties of individual vortex
lines that were imaged and manipulated by magnetic force microscopy%
~\cite{auslaender09}.

Figure~\ref{fig:summary}a depicts the collective weak pinning scenario in
which we are interested.
The red and blue line represents the vortex, with the inner red and outer blue
tubes corresponding to the vortex core and the region of non-zero magnetic
inductance, respectively.
Small grey spheres represent point-like impurities.
The arrows near $H$ and $f_L$ define the directions of the rf magnetic field
and the Lorentz force, respectively.
We also show the depth coordinate $z$ and the displacement field $u(z)$, 
from a reference configuration (dashed line).

The near-depinning behavior of $d$-dimensional manifolds moving in
$d^\prime$-dimensional disordered environments is a long-standing
problem in the field of non-equilibrium statistical mechanics that is
connected to diverse physical situations, from crackling
noise~\cite{sethna01} to raindrops on windshields to superconducting
vortices and plasticity~\cite{fisher98,sethna17}.
In typical vortex pinning models, pinning forces originate in the overlap of
the normal conducting regions associated with the vortex core and the
impurity defect.
Pinning forces associated with atomic impurities are very weak.
Collectively, they add up randomly, so that the average force over a length
$L$ vanishes.
Only fluctuations in either force or impurity density can pin a vortex line.
If the external Lorentz force is small, the vortex line can trade elastic energy
and  find an optimal stationary configuration in the disordered potential
landscape.
Right above the depinning force, the vortex line moves; velocity and
velocity autocorrelations display universal power laws and scaling behavior
associated with emergent scale invariance.
As the Lorentz force increases further away from depinning, the dynamical
behavior crosses over from quenched to dynamic disorder, reminiscent of
the quenched to thermal KPZ crossover~\cite{attis15,kardar86}, and the
vortex line starts moving through unexplored regions of the potential
landscape.

Thus, for collective weak pinning disorder, the vortex line will not move
macroscopically until an external force per unit length becomes greater
than the depinning threshold $f_p$.
The vortex line depinning transition is thought to be continuous --- the
force per unit length resisting the motion of a slowly moving vortex will
approach $f_p$ as the velocity goes to zero (unlike, say, the textbook
behavior of static vs. sliding friction).
Here we shall simulate this depinning explicitly, and also provide a
mean-field model, incorporating the depinning threshold $f_p$ but
ignoring the critical fluctuations, avalanches, and scaling characteristic
of continuous dynamical phase transitions.

Figures~\ref{fig:summary}b and c summarize our main results.
In b, we show a sketch of the behavior of the sensitivity of the residual
resistance to trapped flux as a function of the amplitude of the rf field.
We ignore the regime of very small applied magnetic field also known as
the \emph{Campbell regime}~\cite{campbell71}, in which the vortex
displacements are much smaller than the characteristic pinning length,
the vortex line remains trapped, and the low-dissipation Campbell
response probes the pinning wells~\cite{willa15}.
The sensitivity (black curve) crosses over at $B_{rf}=B_X$ (dashed-green
line) from a linear behavior (red line, with $P\sim {B_{rf}}^3$) at low fields
to a plateau (blue line, with $P\sim {B_{rf}}^2$) at high fields.
Our analysis describes the hysteretic losses dominating the linear
behavior that is observed in the experiments, and the crossover to a
viscous-dominated regime.
In c, we show our calculations for the crossover field $B_X$ (in units of
the thermodynamic critical field $B_c$) as a function of the depinning
current $j_d$ (in units of the depairing current~\cite{blatter94} $j_o$) and
the inverse frequency $f_X / f$, where $f_X$ is a function of superconductor
parameters (see Eq.~\eqref{eq:fXdef}).
We find that $B_X \sim j_d \, f^{-1/2}$.
The blue, green and red lines correspond to the rescaled frequencies of the
Nb$_3$Sn, doped-Nb and NbCu cavities, respectively.

The rest of the paper is organized as follows.
Section~\ref{sec:motion} discusses the vortex equations of motion, and our
solutions for \emph{mean-field} and \emph{local-potential} models based on
collective weak pinning theory.
In Section~\ref{sec:experiments}, we apply our theoretical analysis to new
experimental results for CERN 100MHz NbCu and Cornell 1.3GHz
doped-Nb and Nb$_3$Sn cavities, and discuss possible mechanisms to
justify the high depinning fields that are necessary to explain the
experiments, and the remaining discrepancy between theory and
measurements.
We summarize our results and make some final remarks in
Section~\ref{sec:conclusions}.
In Appendix~\ref{sec:checks}, we present some sanity checks that
corroborate the results presented in Sections~\ref{sec:motion} and%
~\ref{sec:experiments}.
In Appendix~\ref{sec:correction}, we derive the correction factor that we
used in Section~\ref{sec:experiments} to make contact between our
calculations and the experimental measurements.

\section{Vortex motion and dissipation}
\label{sec:motion}

\subsection{Equations of motion}
\label{subsec:equations}

We consider the dynamics of one vortex line in a superconductor that
occupies the half-infinite space ($z>0$).
In its reference configuration, the vortex is a straight line normal to the
the superconductor surface (i.e. the $z=0$ plane).
The vortex configuration at time $t$ is completely determined by the
displacement field $u=u(z;t)$, which in this case is a scalar function of $z$.
Let us write down explicit expressions for some of the terms appearing in
Eq.~\eqref{eq:dynamics1}:
\begin{align}
    f_V
        &= \eta \frac{du}{dt}, \quad
    f_E
        = \epsilon_\ell \frac{d^2 u}{dz^2},
        \nonumber \\
    f_L
        &= \frac{\phi_0 \, H_{rf}}{\lambda} \, e^{-z/\lambda} \sin (2 \pi f \, t),
    \label{eq:forces}
\end{align}
where $\eta$ is the viscosity, $\epsilon_\ell$ is the vortex line tension,
$H_{rf}$ and $\omega$ are the amplitude and frequency of the magnetic
field, $\lambda$ is the superconductor penetration depth, and $f$ is the
rf frequency.
The line tension can be written as~\cite{blatter03,brandt03}
\begin{equation}
	\epsilon_\ell
		= \epsilon_0 \, c(\kappa),
\end{equation}
with
\begin{align}
	& \epsilon_0
		=\phi_0^2 / (4 \pi \mu_0 \lambda^2), \\ \nonumber
	& c \, (\kappa)
		\approx \ln \kappa + 0.5 +\exp(-0.4-0.8\ln \kappa - 0.1 (\ln \kappa)^2),
\end{align}
and $\phi_0$ and $\mu_0$ denoting the fluxoid quantum and the
permeability of free space, respectively.
The viscosity is given by the Bardeen-Stephen formula~\cite{bardeen65}:
$\eta={\phi_0}^2 / (2 \pi \xi^2 \rho_n)$, where $\xi$ is the superconductor
coherence length, and $\rho_n$ is the resistivity of the normal phase.
Defining dimensionless quantities $\tilde{u} = u / \lambda$,
$\tilde{z}=z/\lambda$ and $\tilde{t} = f \, t$, we can combine
Eqs.~\eqref{eq:dynamics1} and \eqref{eq:forces} to write
\begin{equation}
    \frac{d \tilde{u}}{d \tilde{t}}
        = C \left(
            \frac{B_{rf}}{\sqrt{2} B_c} e^{-\tilde{z}} \sin(2 \, \pi \tilde{t})
            + \frac{c \, (\kappa)}{2\kappa} \frac{d^2 \tilde{u}}{d\, \tilde{z}^2}
            +\frac{\xi f_P}{2 \, \epsilon_0}
            \right),
    \label{eq:dynamics3}
\end{equation}
where $C = \rho_n / ( \mu_0 \lambda \, \xi f \kappa^2)$, $B_{rf}$
denotes the amplitude of the rf magnetic inductance, and $B_c$ is the
thermodynamic critical field.

\begin{figure}[!ht]
\includegraphics[width=0.9\linewidth]{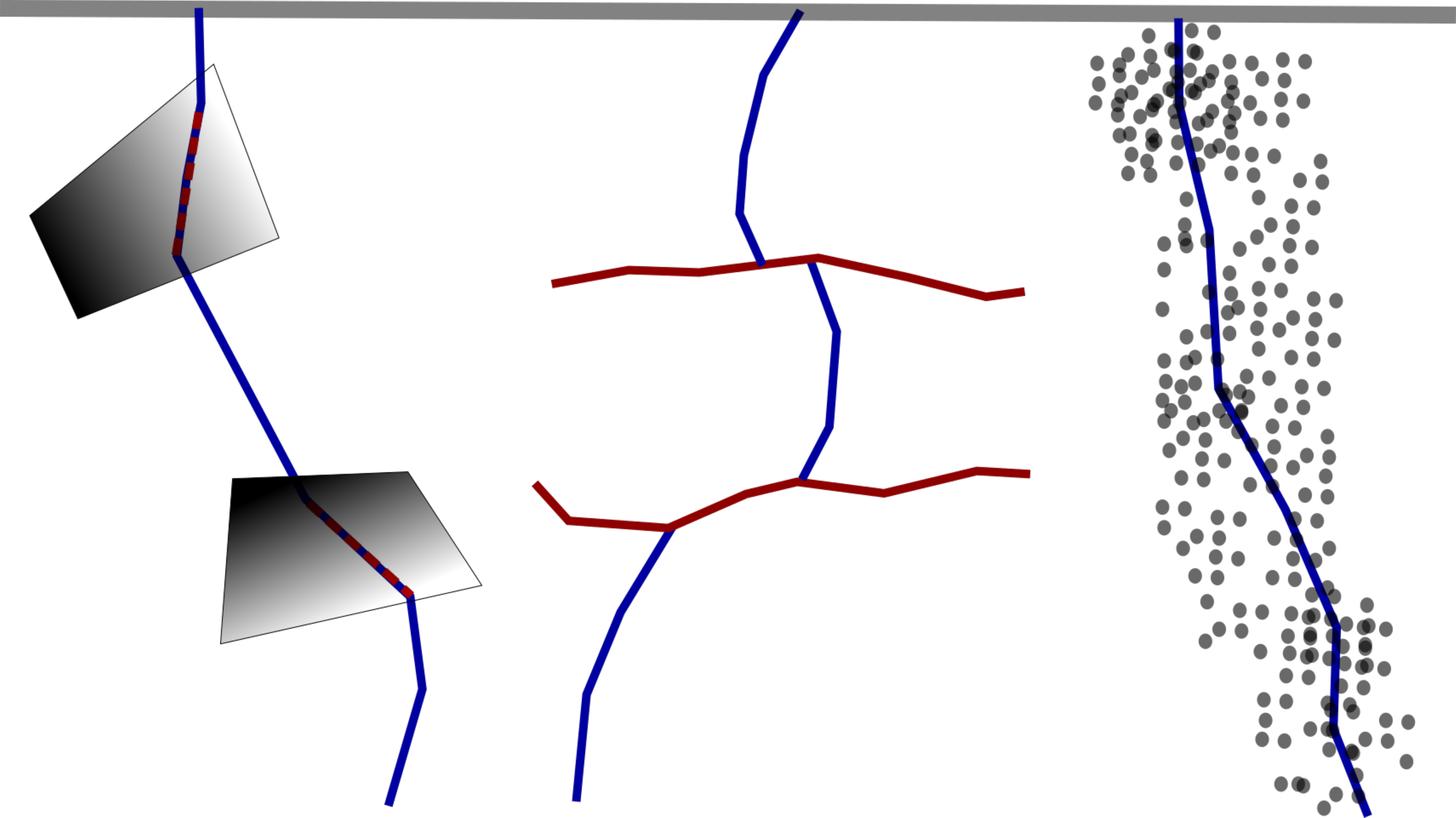}
\caption{
Illustration of the collective weak pinning scenario for general
$\mathcal{D}$-dimensional defects.
The gray line at the top represents the superconductor-vacuum interface.
The blue lines on the left, center and right represent vortex lines.
Small circles on the right correspond to point-like impurities with
$\mathcal{D}=0$ (as in Fig.~\ref{fig:summary}a).
Red lines at the center correspond to line defects with $\mathcal{D}=1$,
such as dislocation lines.
Grey polygons on the left correspond to surface defects with
$\mathcal{D}=2$, such as grain boundaries; the red dashed lines
illustrate the regions where the vortex is pinned by the defects.
\label{fig:defects}}
\end{figure}

In collective weak pinning theory~\cite{larkin79,blatter03}, the
accumulated pinning force over a length $L$ is given by the square-root
fluctuation form,
\begin{equation}
F_P (L) \approx \sqrt{ {F_i}^2 \, n_{\mathcal{D}} \, \xi^{2-{\mathcal{D}}} L }, 
\label{eq:pinningForceL}
\end{equation}
where $F_i$ denotes a typical individual pinning force, $\mathcal{D}$ is the
spatial dimension of the defects (0, 1 and 2 for point-like, line and 
surface defects, respectively), and $n_0$, $n_1$
and $n_2$ are the number of defects per unit volume, area, and length,
respectively%
	\footnote{Note that $n_{\mathcal{D}} \, \xi^{2-{\mathcal{D}}}$ corresponds
	to the  number of individual forces per unit length.}.
Note that standard collective weak pinning theory assumes point-like defects
($\mathcal{D}=0$ in our notation).
For higher-dimensional defects ($\mathcal{D}>0$), we consider a scenario
where the line or surface defects are randomly placed and randomly oriented,
as illustrated in Fig.~\ref{fig:defects}.
The normal-conducting core of the vortex line is attracted to the defect region
and can exhibit pinning and depinning behavior similar to that of point-like
impurities.
Using the superconductor condensation energy, we estimate $F_i$ for
point-like impurities and extended defects such as dislocations and grain
boundaries (see Appendix~\ref{sec:checks}).
Note that pinning by extended defects can be substantially stronger
than pinning by point-like defects.
At lengths larger than the depinning length $L_c$, defined as the length
in which the pinning energy balances the elastic energy, a vortex can
deform and trade elastic energy to find a favorable configuration in the
disordered potential landscape (cutting off the square-root dependence
of the pinning force).
In the standard theory, the vortex line breaks up into a chain of
segments of length $L_c$, each individually competing with the Lorentz
force.
We propose and discuss approximate formulas for the collective pinning
force in Sections~\ref{subsec:mf} and~\ref{subsec:lp}.

The power dissipated by a single oscillating vortex is given by
\begin{align}
    P_1 
        & = f \int_{0}^{1/f} dt \int_0^\infty dz \, f_L \cdot \frac{du}{dt}
            \nonumber \\
        & = \frac{f \lambda \, \phi_0 B_{rf}}{\mu_0}\int_0^1 d\tilde{t} \sin 2 \pi \tilde{t}
            \int_0^\infty d\tilde{z} \, e^{-\tilde{z}} \, \frac{d \tilde{u}}{d \tilde{t}}.
    \label{eq:power}
\end{align}
The net flux trapped in an area $s$ breaks up into $N$ vortices of fluxoid quanta
$\phi_0$, $B_{\text{trap}} s = N \phi_0$, so that, using Eq.~\eqref{eq:surfaceR}
to calculate the residual resistance, we find:
\begin{align}
    \frac{R_0}{B_{\text{trap}}}
        & = \frac{2 \, {\mu_0}^2 P_1}{\phi_0 {B_{rf}}^2}
            \nonumber \\
        & = \frac{2 f \lambda \, \mu_0}{B_{rf}} \int_0^1 d\tilde{t} \sin 2 \pi \tilde{t}
            \int_0^\infty d\tilde{z} \, e^{-\tilde{z}} \, \frac{d \tilde{u}}{d \tilde{t}}.
    \label{eq:sensitivity}
\end{align}

\subsection{Mean-field model}
\label{subsec:mf}

In this section, we consider a mean-field version of the pinning force using the
collective weak pinning theory. 
We assume that the absolute value of the pinning force is the \emph{depinning}
force, i.e. the Lorentz force due to a transverse uniform current accumulated over
the depinning length $L_c$, and that its sign is chosen so that it opposes the sum
of the Lorentz and the elastic forces,
\begin{equation}
    f_P = - \text{sgn}(f_L + f_E) \, \phi_0 \, j_d,
    \label{eq:mf_pinning}
\end{equation}
where sgn denotes the sign function and $j_d$ is the depinning
current.
Equation~\eqref{eq:mf_pinning} is a key assumption on our mean-field model,
and partly follows from the force balance equation~\eqref{eq:dynamics1}.
If the frequency is small, we can ignore the viscous dissipation force
in~\eqref{eq:dynamics1}, which leads to a pinning force that opposes the sum
of the elastic and Lorentz forces, thus justifying the sign function.
The constant appearing in Eq.~\eqref{eq:mf_pinning} also follows from the force
balance equation~\eqref{eq:dynamics1}, and collective weak pinning theory.
If the motion is quasi-static, at each time the vortex line accommodates itself in
the rugged potential landscape to minimize its free energy, deforming over lengths
of order the depinning length $L_c$.
As previously mentioned, we can break up the vortex line into smaller segments
of size $L_c$ and assume that the pinning force balances the Lorentz force for
each segment.
The segments will not move away from their low-energy configuration until the
Lorentz force overcomes the pinning force; so we assume the pinning force is given
by the Lorentz force ($\phi_0 j$) at the ``critical'' depinning current $j = j_d$, which
is a convenient and experimentally measurable quantity)%
\footnote{
The idea of a ``critical'' force also appears in a critical state
model~\cite{tinkham96,huebener01}, such as the Bean model~\cite{bean62}, but in
a different context.
However, unlike the Bean model, our model ignores the interactions between
vortices and incorporates the structure of the vortex line.
The Bean model involves many interacting vortices pinned on dirt; our model is a
single vortex pinned (collectively) on many dirt particles.}.
Note that $f_P$ is a piecewise function, with each sub-domain being determined
by the sign of $f_L+f_E$, instead of the expected depinning length $L_c$.
This simplifying assumption allows us to gain insight from approximate analytical
solutions, and is motivated by the fact that we consider ranges of large magnetic
fields, far above depinning, so that we expect the realistic model to display
fairly smooth solutions.
We show in Section~\ref{subsec:lp} that our numerical simulations of the local
potential model corroborate this assumption.

First we consider the low frequency behavior, where the vortex motion is slow,
and we can neglect the viscous dissipation term.%
  \footnote{One must note that the low-frequency limit approaches the 
  depinning transition, where disorder-induced fluctuations become important
  and the mean-field model is not quantitatively correct. It is, however,
  analytically solvable and a useful illustration and starting point for
  understanding high-frequencies (Section~\ref{sec:experiments}) and 
  interpreting the local potential simulations incorporating disorder
  (Section~\ref{subsec:lp}).}
This approximation is valid for the range of parameters in which\
$\eta \, v_\text{max} / |f_P| \ll 1$, where $v_\text{max}$ is the maximum velocity
of the vortex displacement field at the boundary.
We revisit this condition later on this section, when we self-consistently define
the crossover from cubic to quadratic dissipation.
We also make a point-force approximation, by replacing the exponential decay
of the Lorentz force by a delta function:
$\exp(-\tilde{z}) \approx \delta (\tilde{z})$.
This approximation is adequate when the amplitude of motion in the $z$
direction ($a_z$) is sufficiently large compared to the penetration depth
$\lambda$.
Note that the existence of a delta function at the boundary fixes the slope of
the displacement at $z=0$ for each time, violating, in general, the realistic
constraint of zero normal current at the superconductor surface
($d \tilde{u} / d\tilde{z} = 0$ at $\tilde{z}=0$).%
	\footnote{In Appendix~\ref{sec:checks}, we deform our analytical solution
	over a length $\lambda$ near the boundary to satisfy the constraint at the
	surface.	
	For large enough fields (in particular, for most of the range of fields
	considered in Fig.~\ref{fig:sensitivity}),the change in vortex length is very
	small compared to the amplitude of motion in the $y$ direction, suggesting
	that the error resulting from this approximation is small.}
Now equation~\eqref{eq:dynamics3} can be written as
\begin{equation}
    \frac{d^2 \tilde{u}}{d \tilde{z}^2}
        = \pm \alpha - \beta \, \sin (2\, \pi \tilde{t}) \delta (\tilde{z}), 
    \label{eq:dynamics4}
\end{equation}
where the $\pm$ depends on the value of the sgn function in 
Eq.~\ref{eq:mf_pinning}, and where $\alpha$ and $\beta$ are given by
\begin{equation}
    \alpha
        = \frac{\lambda |f_P|}{\epsilon_\ell}, \quad
    \beta
        = \frac{\sqrt{2} \, \kappa}{c(\kappa)} \frac{B_{rf}}{B_c}.
\end{equation}

The solution of Eq.~\eqref{eq:dynamics4} is a parabola:
\begin{equation}
    \tilde{u} (\tilde{z})
        = a + b \, \tilde{z} \pm \frac{\alpha}{2} {\tilde{z}}^2.
    \label{eq:mf_solution1}
\end{equation}
where $a$ and $b$ are constants determined by the boundary conditions.
Integration of Eq.~\eqref{eq:dynamics4} over a small interval near the surface
leads to
\begin{equation}
    \left. \frac{d\tilde{u}}{d\tilde{z}} \right|_{\tilde{z} = 0^+}
        = b
        = -\beta \, \sin (2\pi \tilde{ t }),
\label{eq:BC1}
\end{equation}
and,
\begin{equation}
    \tilde{u} (\tilde{z})
        = a -\beta \, \sin (2\pi \tilde{t}) \, \tilde{z} \pm \frac{\alpha}{2} {\tilde{z}}^2.
    \label{eq:mf_solution2}
\end{equation}
Equation~\eqref{eq:mf_solution2} is only valid at sufficiently small $z$; the
vortex line remains pinned in the superconductor deep interior.
We find $a$ by imposing that the vortex moving section continuously and
smoothly merges with the pinned section at a distance $\tilde{z}^*$ that we
determine.
Let $\tilde{u}_<$ and $\tilde{u}_>$ be the solutions near and away from the
superconductor surface, respectively. The complete solution is given by
\begin{equation}
    \tilde{u}
        = \begin{cases}
            \tilde{u}_<, & \text{for } \tilde{z} < \tilde{z}^*, \\
            \tilde{u}_>, & \text{otherwise,}
        \end{cases}
\end{equation}
where $a$ and $\tilde{z}^*$ are determined by the equations:
\begin{equation}
    \tilde{u}_< (\tilde{z}^*)
        = \tilde{u}_> (\tilde{z}^*), \quad
    \frac{d \tilde{u}_<}{d \tilde{z}} (\tilde{z}^*)
        = \frac{d \tilde{u}_>}{ d \tilde{z}} (\tilde{z}^*). 
\label{eq:vsaEqs}
\end{equation} 

\begin{figure}[!ht]
\includegraphics[width=0.8\linewidth]{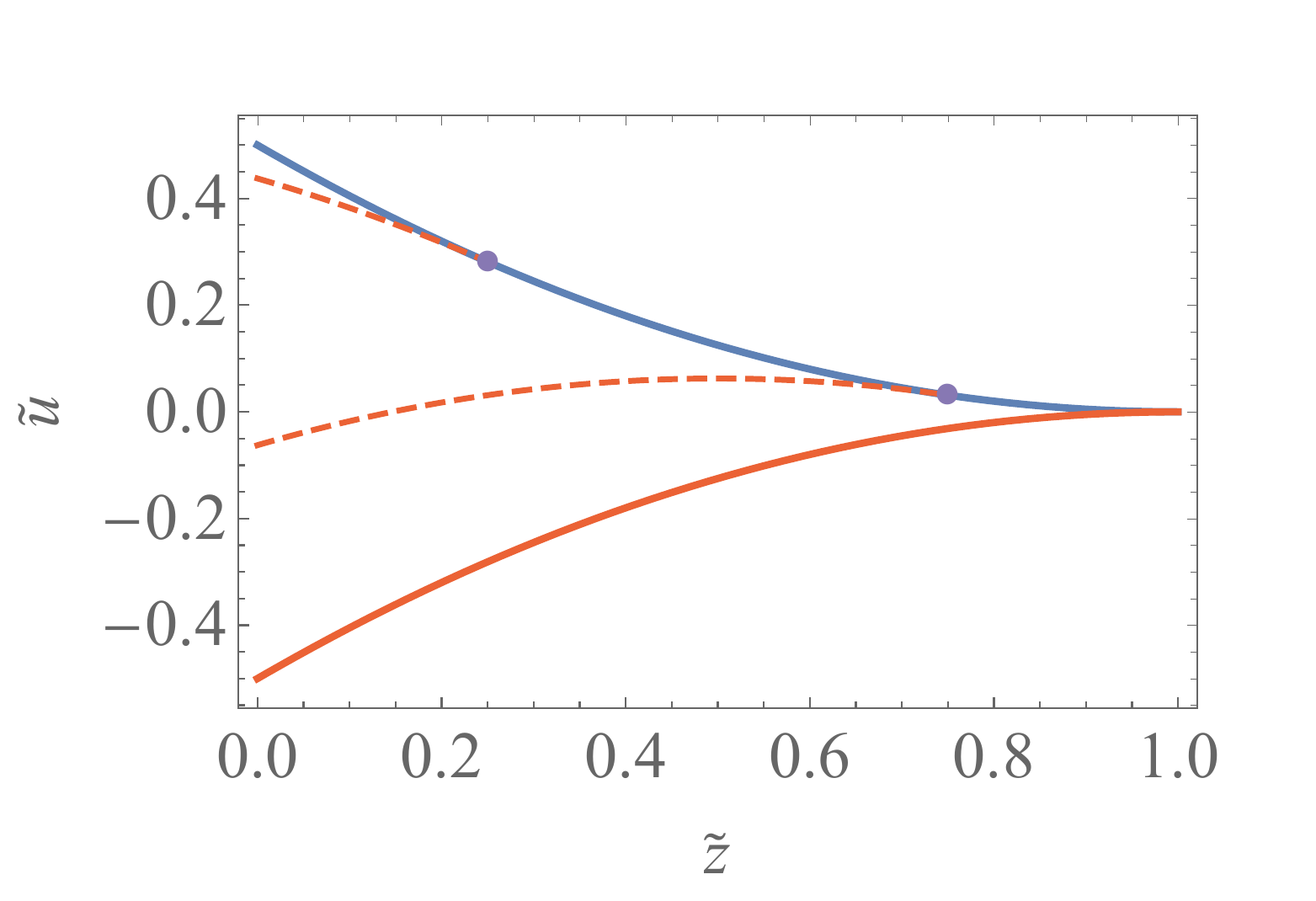}
\caption{
Mean-field solution of the vortex displacement field $\tilde{u}$ as a function
of depth coordinate $\tilde{z}$ for $\tilde{t} = 1/4$ (blue), $5/12$ and
$7/12$ (dashed red) and $3/4$ (solid red).
\label{fig:uSolutions}}
\end{figure}

Let us study the solutions for $\tilde{t} \in [0,1/4]$, assuming
$\tilde{u}(\tilde{z};\tilde{t}=0)=0$.
We use the subscript $0$ to denote solutions in this interval.
Using Eqs.~\eqref{eq:mf_solution2} and~\eqref{eq:vsaEqs}, we find,
\begin{equation}
    \tilde{u}_0 (\tilde{z}; \tilde{t})
        = \begin{cases}
            \dfrac{\alpha}{2} \left(\tilde{z} - \dfrac{\beta}{\alpha}
                \sin (2\pi \tilde{t}) \right)^2,
            & \text{for } \tilde{z} < \dfrac{\beta}{\alpha} \sin (2\pi \tilde{t}), \\
            0,
            & \text{otherwise.} 
\end{cases}
\end{equation} 
The blue line in Fig.~\ref{fig:uSolutions} corresponds to $\tilde{u}_0$ as a
function of $\tilde{z}$ for $\tilde{t} = 1/4$ and $\alpha=\beta=1$.
As $\tilde{t}$ increases from $1/4$, the elastic and pinning forces
exchange signs near the surface, the tip of vortex line reverses motion and
starts ``unzipping'' from the the blue curve.
The complete solution has $\tilde{u}_>=\tilde{u}_0 (\tilde{z};1/4)$ and
$\tilde{u}_<$ given by~\eqref{eq:mf_solution2} with the negative sign (red
curves in Fig~\ref{fig:uSolutions}), and with $a$ and $\tilde{z}^*$ satisfying
Eq.~\eqref{eq:vsaEqs}.
For $\tilde{t} \in [1/4,3/4]$, we find,
\begin{equation}
    \tilde{u}(\tilde{z}; \tilde{t})
        = \begin{cases}
                a(\tilde{t}) - \beta \sin (2\pi \tilde{t}) \tilde{z} - \dfrac{\alpha}{2}
                    \, {\tilde{z}}^2,
                & \text{for } \tilde{z} < \tilde{z}^* (\tilde{t}), \\
                \dfrac{\alpha}{2} \left(\tilde{z} - \dfrac{\beta}{\alpha} \right)^2,
                & \text{otherwise.}
        \end{cases}
    \label{eq:uSol}
\end{equation} 
 with
\begin{align}
    & a(\tilde{t})
        = \frac{\beta^2}{8 \alpha} \left(1 + \cos 4 \pi \tilde{t}
            + 4 \sin 2 \pi \tilde{t} \right), \\
    & \tilde{z}^* (\tilde{t})
        = \frac{\beta}{ 2 \alpha} (1 - \sin 2 \pi \tilde{t} ).
\end{align}

Note that the amplitude of motion at the surface is proportional to
$\tilde{u}(0,1/4) \propto \beta^2 \propto {B_{\text{rf}}}^2$, so that the
dissipation energy is proportional to
$f_L \times {B_\text{rf}}^2 \propto {B_\text{rf}}^3$, in agreement with the
experiments.
This leads to the important conclusion that the cubic dissipation is
intimately connected to the quadratic solutions for the vortex motion,
which is an ultimate consequence of the existence of a pinning force
$\alpha$.
One caveat: The cubic dissipation might become quadratic when the
boundary condition in the deep interior of the superconductor is changed.
For instance, a simple way of controlling the total dissipation consists in
employing restrictive \emph{inescapable} pinning potentials (such as the
ones considered in references~\cite{checchin17} and~\cite{gurevich13})
for the vortex line at a distance $\tilde{z}_p$ so that
$\tilde{u}(\tilde{z}_p) \approx 0$.
Our simple calculations show that if $\tilde{z}_p$ is sufficiently small (for
a given field), the dissipation is proportional to ${B_{\text{rf}}}^2$; the
cubic behavior disappears.
In Section~\ref{subsec:mechanisms} we discuss how the combination of
strong and collective weak pinning might help explain the discrepancy
between theory and experiments.

Figure~\ref{fig:uSolutions} shows solutions of $\tilde{u}$ as a function of
$\tilde{z}$, for $\alpha=\beta=1$, and $\tilde{t}=1/4$ (blue), $5/12$
and $7/12$ (dashed red), and $3/4$ (solid red).
The purple dots correspond to the points where the two parabolas merge.
The subsequent solution in the interval $[3/4, 5/4]$, is a reflection of the
solutions in $[1/4, 3/4]$, i.e. $\tilde{u}(\tilde{t}) = -\tilde{u}(\tilde{t}-1/2)$, for
$\tilde{t} \in [3/4, 5/4]$.

We use Eq.~\eqref{eq:power} to write down the power dissipated by one
vortex,
\begin{equation}
	P_1
		= \frac{8 \pi}{3 \, {\mu_0}^2} \frac{f \, \lambda^2}{j_d \, c(\kappa)}
			{B_\text{rf}}^3,
\end{equation}
and Eq.~\eqref{eq:sensitivity} to calculate the sensitivity of the residual
resistance to trapped flux,
\begin{equation}
    \frac{R_0}{B_{\text{trap}}}
        = A B_{rf},
    \label{eq:sensitivity_analytical}
\end{equation}
where
\begin{equation}
	A
		= \frac{16 \pi}{3\, \phi_0} \frac{f \, \lambda^2}{c(\kappa) j_d}.
	\label{eq:sensitivity_slope}
\end{equation}
The sensitivity linear increase with the rf field is qualitatively consistent
with experimental measurements for 100MHz NbCu and 1.3GHz
doped-Nb and Nb$_3$Sn cavities (see Section~\ref{sec:experiments}).
For better \emph{quantitative} agreement with the experimental results,
we include a correction to account for the alignment of the vortices
throughout the cavity surface and field depletion at the cavity poles (see
Appendix~\ref{sec:correction}).
Note that the measured residual resistance approaches a finite value as
$B_\text{rf} \rightarrow 0$, whereas our pinning model predicts
$R_0 \rightarrow 0$ at the same limit
(Eq.~\eqref{eq:sensitivity_analytical}).
We ignore other sources of residual resistance that are not associated
with vortex motion, and that can explain this offset.
For example, a plausible model of \emph{static} residual resistance
considers the normal conducting resistance originating in the core of the
vortex line~\cite{padamsee08}.
To make a direct comparison with experiments, we subtract off the offset
of the measured sensitivity (red circles) in Fig.~\ref{fig:sensitivity}.
In these modern cavities, the linear term we attribute to vortex motion
dominates $R_0$ under operating conditions; the offset is equal to $7.7$,
$4.6$ and $0.03$ n$\Omega/\mu$T in Fig.~\ref{fig:sensitivity}, for doped
Nb, Nb$_3$Sn and NbCu, respectively.

The hysteretic losses that are responsible for the linear slope of the
sensitivity become less important at high rf field amplitudes, so that we
expect a crossover to a high-fields regime where viscous dissipation is
the dominant loss mechanism.
To quantify this crossover, we use the solution given by
Eq.~\eqref{eq:uSol} to self-consistently calculate $\eta \, v_\text{max}$,
which we compare with the pinning force.
Here $v_\text{max}$ is the maximum of $du / dt$ at $z=0$ over one
period of oscillation.
We define the crossover field $B_X$ from the equation:
\begin{equation}
    \left. \frac{\eta \, v_\text{max}}{f_P} \right|_{B_{rf} =B_X}
        = 1,
\end{equation}
yielding,
\begin{equation}
	B_X 
		= \sqrt{\frac{2}{3\sqrt{3}\pi} \frac{\mu_0 \, \rho_n \, c\,(\kappa) \,
			{j_d}^2}{\kappa^2 f} }.
	\label{eq:crossoverB}
\end{equation}
Equation~\eqref{eq:crossoverB} can also be written in the dimensionless
form
\begin{equation}
	\frac{B_X}{B_c}
		= \sqrt{\frac{f_X}{f}} \frac{j_d}{j_o},
\end{equation}
where,
\begin{equation}
	f_X
		\equiv \frac{16}{81\sqrt{3}\pi}
			\frac{c \, (\kappa)}{\kappa^2}
			\frac{\rho_n}{\mu_0 \, \lambda^2},
	\label{eq:fXdef}
\end{equation}
and the depairing current is given by
\begin{equation}
	j_o
		= \frac{4}{3\sqrt{6}} \frac{B_c}{\mu_0 \lambda},
	\label{eq:depairingGL}
\end{equation}
according to Ginzburg-Landau (GL) theory.
We have already briefly discussed Figure~\ref{fig:summary}c, showing
the crossover field $B_X/B_c$ as a function of the depinning current
$j_d/j_o$ and the inverse frequency $f_X/f$, with blue, green and red
lines corresponding to the Nb$_3$Sn, doped-Nb and NbCu cavity
rescaled frequencies, respectively.
Table~\ref{tab:parameters} show our calculated values for $B_X$ using
the simulation parameters.
Note that low-$\kappa$, low-frequency SRF cavities at high depinning
currents have high $B_X$.

The viscous dissipation term is important and cannot be neglected at
either high frequency or high field amplitudes.
Finding closed forms for the piecewise solutions of the full equation of
motion is beyond the scope of this paper.
We then opted for discretizing the vortex line using Python arrays, and
using SciPy odeint package to numerically integrate
Eq.~\eqref{eq:dynamics3}.
We give more details of these simulations in
Section~\ref{sec:experiments}.

\subsection{Local-potential model}
\label{subsec:lp}

\begin{figure}[!ht]
\begin{minipage}[t]{0.67\linewidth}
\centering (a) \par\smallskip
\includegraphics[width=\linewidth]{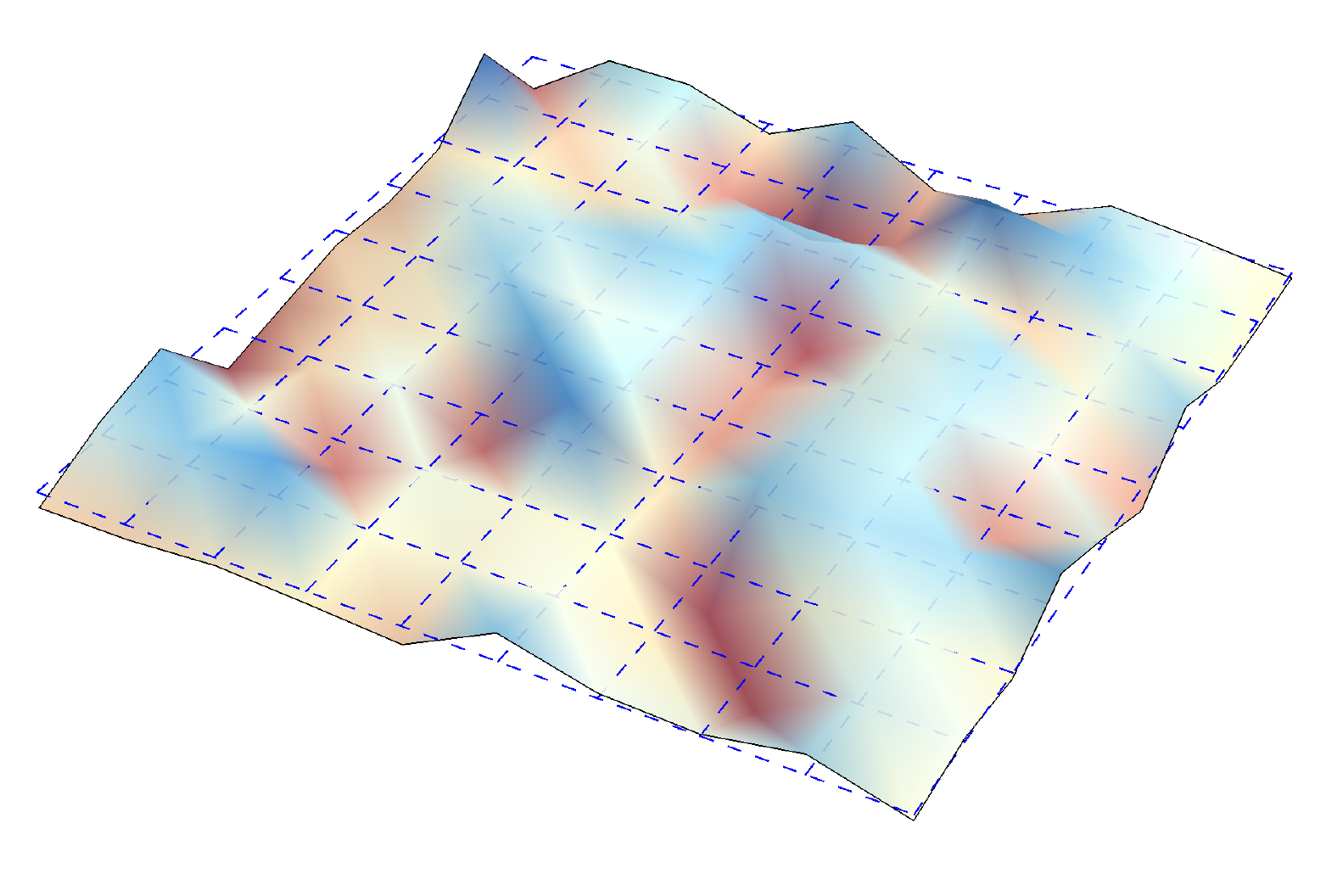}
\end{minipage}
\hspace{0.01cm}
\begin{minipage}[t]{0.29\linewidth}
\centering (b) \par\smallskip
\includegraphics[width=\linewidth]{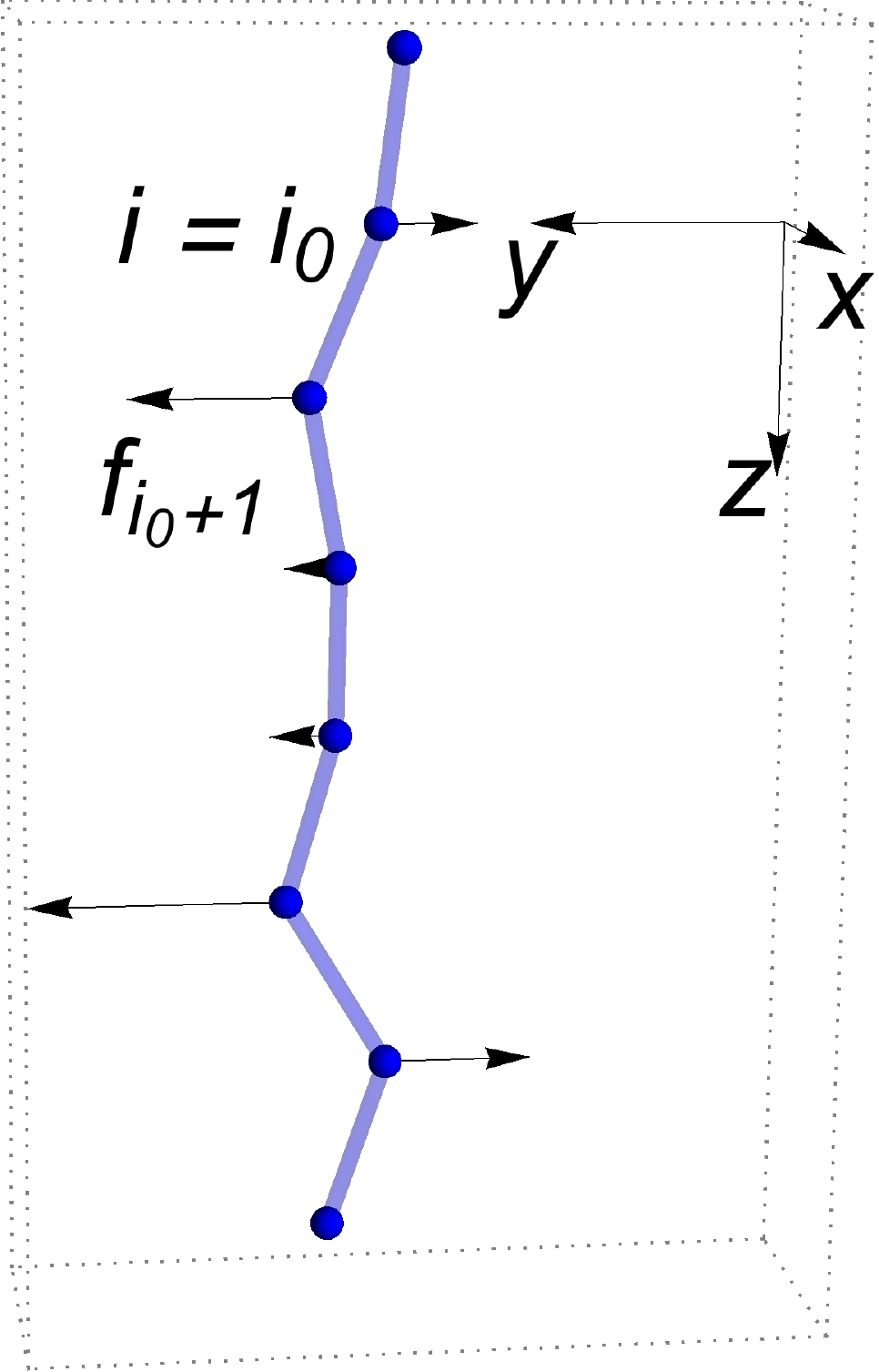}
\end{minipage}
\caption{%
(a) Illustration of a disordered potential landscape in the collective
weak pinning scenario.
(b) Discretized vortex line (blue dots and segments), subject to local
random forces (black arrows) originating in the disordered potential
landscape shown in (a).
\label{fig:local_potential}}
\end{figure}

Here we consider a model where the vortex line is subject to local
pinning forces originating in a Gaussian random potential of zero mean,
and adequately scaled variance.
In our numerical approach
we first define a grid of spacing $\tilde{a}<a$ in the $y$-$z$ plane, where
$a$ is the spacing of the $z$-coordinate of the vortex array (we choose
$\tilde{a} \sim \xi$, which is the smallest length a superconductor can
resolve).
We then assign i.i.d. normal random variables to each point of the grid,
and use a spline interpolant to
implement the unscaled potential $\tilde{U}$ for
arbitrary $y$ and $z$.
Figure~\ref{fig:local_potential}a depicts a square grid (blue-dashed lines)
and the corresponding interpolated potential.
The individual force per length acting on a segment $i$ of the discretized
vortex is then given by $f_i= (U_0/a) \, \partial \tilde{U} / \partial y$, where
the constant $U_0$ is chosen to match the depinning force determined by
collective weak pinning theory.
The force accumulated over the depinning length $L_c$ can be written
as
\begin{equation}
    F_P (L_c)
        = \sqrt{ \left< \left(  \sum_{i=i_0}^{i_0+L_c/a} f_i \, a \right)^2\right>},
\end{equation}
where $i_0$ is an arbitrary initial site of the vortex array (see
Fig.~\ref{fig:local_potential}b), and $\langle \cdot \rangle$ denotes an
average over $i_0$.
We expect local individual forces be uncorrelated over distances
$\gtrsim 2 \, \tilde{a}$, so that, after some algebra,
\begin{equation}
    F_P (L_c)
        \approx U_0 \sqrt{a \, L_c} \, \sigma_{\tilde{f}},
    \label{eq:interpolatedPinning}
\end{equation}
where $\sigma_{\tilde{f}}$ is the variance of $\tilde{f}_i \equiv f_i / U_0$.
Equations~\eqref{eq:interpolatedPinning} and~\eqref{eq:pinningForceL},
result in
\begin{equation}
	U_0
		= \frac{\phi_0 \, j_d}{\sigma_{\tilde{f}}} \sqrt{ \frac{L_c}{a} }.
\end{equation}
We use collective weak pinning theory to express $L_c$ as a function
of $j_d$.
Let $\gamma \equiv  {F_i}^2 \, n_{\mathcal{D}} \, \xi^{2-{\mathcal{D}}}$ in
Eq.~\eqref{eq:pinningForceL}, so that $f_P = \sqrt{\gamma/L}$ is the
pinning force per length $L$.
The pinning energy per length is then given by
$\xi f_P = \sqrt{\gamma \, \xi^2 / L}$.
To find the depinning length, we minimize the total energy per length
with respect to $L$ for a small displacement (of order $\xi$) of the
vortex line in the absence of the Lorentz force, i.e.,
\begin{equation}
	\frac{d}{d \, L} \left[ \frac{\epsilon_\ell}{2} \left(\frac{\xi}{L}\right)^2
		- \sqrt{ \frac{\gamma \, \xi^2}{L}} \right]_{L=L_c}
		=0,
\end{equation}
resulting in
\begin{equation}
	{L_c}^3
		= \frac{4 \, {\epsilon_\ell}^2 \, \xi^2}{\gamma}
		= \frac{4 \, {\epsilon_\ell}^2 \, \xi^{\mathcal{D}}}{n_{\mathcal{D}} {F_i}^2}.
	\label{eq:LcCube}
\end{equation}
Now we make $f_P$ equal the Lorentz force due to a transverse
uniform current $j_d$,
\begin{equation}
	f_P
		= \sqrt{\frac{\gamma}{L_c}}
		= \phi_0 \, j_d,
\end{equation}
to eliminate $\gamma$ in~\eqref{eq:LcCube}, finding,
\begin{equation}
	{L_c}^2
		= \frac{2 \, \epsilon_\ell \, \xi}{\phi_0 \, j_d}
		=  \frac{\sqrt{2} \, B_c \, c \, (\kappa) / (\lambda \, \mu_0)}{j_d}
			\, {\xi}^2.
	\label{eq:LcSq}
\end{equation}
Equation~\eqref{eq:LcSq} is usually written in the approximate
form~\cite{blatter94}: $L_c/\xi \approx \sqrt{ j_o / j_d}$, where $j_o$ is
the depairing current calculated using GL theory (see
Eq.~\eqref{eq:depairingGL}).
Collective weak pinning is valid when $L_c \gg \xi$, or $j_d \ll j_o$.
We present our simulation results for doped Nb, Nb$_3$Sn and NbCu
along with the experimental results of Section~\ref{sec:experiments}.

\section{Experiments and simulations}
\label{sec:experiments}

In this section, we discuss our numerical simulations, and make
contact with experimental measurements performed in CERN and
Cornell.
In Section~\ref{subsec:setup} we discuss the experimental setup
for doped Nb, Nb$_3$Sn and NbCu cavities.
In Section~\ref{subsec:simulations} we give additional details of
the simulations, and present experimental, analytical and
simulation results for the sensitivity to trapped flux of the residual
resistance.
In Section~\ref{subsec:mechanisms} we discuss plausible mechanisms
that might explain the discrepancy between theory and experiments.

\subsection{Experimental setup}
\label{subsec:setup}

\subsubsection{Doped Nb}
\label{subsubsec:dopedNb}

Niobium cavities impurity doped with nitrogen in a high-temperature
furnace show a characteristic field-dependent decrease in the BCS
surface resistance that is frequently referred to as
``anti-Q-slope''~\cite{grassellino13}.
In the last few years, significant effort has gone into the study of the
science of impurity doped niobium (see for
example~\cite{gonnella16b,maniscalco17,martinello17}), and nitrogen
doped 1.3 GHz SRF cavities have now found their first use in the
LCLS-II accelerator~\cite{liepe14}.
The residual resistance of nitrogen doped niobium cavities due to
trapped flux has been shown to strongly depend on the electronic
mean free path of the niobium in the rf penetration layer, with a
characteristic bell shape dependence of
$R_0$~\cite{gonnella16,checchin17}.
Recent results indicate that the anti-Q slope is not unique to nitrogen
doping, but can also be found in higher frequency (multi-GHz) SRF
cavities without doping~\cite{martinello17,koufalis18}, as well as in
1.3 GHz cavities with high concentrations of oxygen and carbon
dissolved in the surface~\cite{koufalis18b}.
As part of our studies on the field dependence of the trapped flux
residual resistance, we measured trapped flux losses in a 1.3 GHz
cavity that had been heat treated at 160C for 48 hr in an Ar/CO$_2$
gas mixture (99.99999\% purity Ar gas mixed with 10 ppm CO$_2$)
immediately following an 800C vacuum anneal and prior to rf
performance testing.
Secondary ion mass spectroscopy (SIMS) analysis of a witness
sample revealed very high concentrations of C and O---especially
within the first few 100 nm~\cite{koufalis18b}.
 
The performance of the impurity doped cavity and its sensitivity to
trapped magnetic flux was measured in a standard SRF vertical test
setup, with a uniform ($\pm$10\%), ambient DC magnetic field applied
along the direction of the cavity axis by a Helmholtz-coil during
cool-down (refer to~\cite{gonnella16} for details on this setup).
Using standard cavity rf measurement techniques, the quality factor
of the cavity was measured as function of rf field amplitude,
temperature, and trapped magnetic field, from which the average
additional surface resistance caused by the trapped flux was estimated
as a function of the strength of the rf field.
The results showed a clear linear dependence on the rf field, i.e. a cubic
field dependence of the trapped vortex losses.

\subsubsection{Nb$_3$Sn}
\label{subsubsec:Nb3Sn}

The A15 superconductor Nb$_3$Sn is a particularly promising material for
next-generation, high-performance SRF cavities~\cite{posen17}.
Cornell University has a leading Nb$_3$Sn SRF research program that aims
at exploring the full potential of this material~\cite{posen14}.
Nb$_3$Sn coatings of a few microns thickness are produced on Nb substrate
cavities via a tin vapor diffusion process~\cite{mueller96,posen11}.
Optimization of this process has resulted in the first Nb$_3$Sn SRF
accelerator cavities ever to clearly outperform traditional solid-niobium
cavities in cryogenic efficiency at usable accelerating fields.
Cornell's 1.3GHz Nb$_3$Sn cavities are now routinely reaching quality
factors at 4.2K in the 1 to $2 \times10^{10}$ range~\cite{posen15b}, more
than one order of magnitude above those reachable with niobium at that
temperature and rf frequency.
Due to the bi-metal structure of these cavities, very small spatial thermal
gradients are essential during cool-down to minimize thermoelectrically
induced magnetic fields, that could be trapped and cause significant losses
in rf fields.
However, because of the small thermal gradients during cool-down,
expulsion of residual ambient magnetic fields is poor, therefore still resulting
in some trapped magnetic flux.
Understanding the sensitivity of the residual resistance to trapped flux is
therefore of particular importance for Nb$_3$Sn cavities. 

We used the same experimental procedure discussed in
Section~\ref{subsubsec:dopedNb} for doped Nb.
The results also showed a clear linear dependence on the rf field.

\subsubsection{NbCu}
\label{subsubsec:NbCu}

Quarter-wave resonators of Nb films have been developed for the
post-acceleration of heavy ions at CERN (HIE-ISOLDE
project~\cite{kadi17}).
The Nb film of a few microns is deposited on a Cu cavity by the DC-bias
sputtering technique.
The resonant frequency is 101.28~MHz and operation temperature is
4.5~K.
As the cavity is made of the thin film, its crystal structure contains fine
grains and dislocations inside~\cite{sublet14}.
Flux expulsion during cooling down is typically poor because of a lot of
possible pinning centers, uniformity of temperature caused by the Cu
substrate, and QWR geometry.
Hence, an ambient field can be fully trapped by the cavity.
The bi-metal structure also gives rise to a possible thermoelectrically
induced magnetic field during the cooling down, to be trapped by the
pining centers~\cite{miyazaki_SRF2017}.

The performance of the cavity was evaluated by the standard rf
measurement and magnetometry of representative
samples~\cite{miyazaki18}.
In the rf measurement, the quality factor of the cavity was obtained
by field-decay and coupling information.
From the quality factor along with a geometrical factor evaluated
by rf simulation, rf surface resistance averaged over the cavity
surface was estimated as a function of the strength of rf fields.
At 2.4~K, where the effect of quasi-particles are negligible i.e. no
BCS resistance, the surface resistance turned out to be linearly
dependent on the rf fields~\cite{miyazaki_thinfilm18}.
This behavior was previously reported in
references~\cite{weingarten95,benvenuti99,ciovati07}.
The magnetometry revealed the depinning current of the Nb film
showing such surface resistance~\cite{miyazaki_TTCFermi_17}.
This de-pinning current is larger than the literature value of clean
bulk Nb, but still well below the surface current caused by the rf
fields.

\subsection{Simulations}
\label{subsec:simulations}

We model the vortex line as a discrete one-dimensional Python
array of size $L$ and spacing $a$.
We use $a=38$nm, $a=13$nm and $a=40$nm in the doped Nb,
Nb$_3$Sn and NbCu simulations, respectively, and $L=128$ for
all simulation data presented in this paper.
Table~\ref{tab:parameters} summarizes the material parameters
used in the simulations.
For each simulation, we start with a straight line vortex,
$\tilde{u}(\tilde{z};0)=0, \,\, \forall z$, and 
and find the solution at a later time $t$ by implementing the equations
as an ordinary differential equation (ODE).
For the mean-field model, we integrate Eq.~\eqref{eq:dynamics1} with
the pinning force given by Eq.~\eqref{eq:mf_pinning} for three cycles
(i.e. three periods of oscillation of the applied magnetic field) to relax
the vortex, and then run the simulation for one additional cycle to
calculate the resistance from an average of the dissipated power.
For the local potential model, we integrate Eq.~\eqref{eq:dynamics1}
for the elastic vortex line moving in the random potential as described
in Section~\ref{subsec:lp}, and use three cycles to relax the vortex
and three cycles to measure the dissipation; we repeat this protocol
for ten random initial configurations of the disordered potential, and
calculate the average%
\footnote{We do not need average over samples in the mean-field
model, which is deterministic.}.
Increasing the number of cycles in the relaxation and measurement
processes does not lead to significant changes.
\renewcommand{\arraystretch}{1.5}
\begin{table*}[!ht]
\begin{center}
\begin{tabular}{| c | r | r | r | r | r | r | r | r |}
\hline
Material & $\lambda$[nm] & $\xi$[nm] & $\kappa$ & $B_c$[mT] &
$\rho_n$[$\Omega \cdot $m] & $j_d$[A/m$^2$] & $f$ [GHz] & $B_X$[mT]
\\ \hline\hline
Doped Nb & 39~\cite{maxfield65} & 38~\cite{maxfield65} & 1 & 152	 &
$1.8 \times 10^{-8}$~\cite{goodman68} & $10^{10}$ & 1.3 & 16
\\ \hline
Nb$_3$Sn & 111~\cite{tinkham96} & 4.2~\cite{orlando79} & 26.4 &
483	& $10^{-6}$~\cite{godeke06} & $10^{10}$ & 1.3 & 8
\\ \hline
NbCu & 30~\cite{miyazaki18} & 30~\cite{miyazaki18} & 1 & 250 &
$4.5\times10^{-9}$	 & $10^{10}$~\cite{miyazaki18,miyazaki_TTCFermi_17}
& 0.1 & 28
\\ \hline
\end{tabular}
\caption{
Penetration depth $\lambda$, coherence length $\xi$,
Ginzburg-Landau parameter $\kappa \equiv \lambda / \xi$,
thermodynamic critical field (according to GL theory)
$B_c = \phi_0 / (2 \sqrt{2} \pi \lambda \xi)$, normal state resistivity
$\rho_n$, depinning current $j_d$, and frequency $f$ used in simulations
for doped Nb, Nb$_3$Sn and NbCu.
The last column shows the crossover field $B_X$, according to
Eq.~\eqref{eq:crossoverB}.
For Nb$_3$Sn, we have used values for $j_d$ that are higher than
reported measurements~\cite{sumption12} for tube-type Nb$_3$Sn
superconductors.
In Appendix~\ref{sec:checks}, we do a sanity check of this higher
threshold, calculating the pinning per impurity assuming each destroys
superconductivity over some region. 
For a mean-free-path of $\sim1$nm, a 1\% density of impurities destroying
superconductivity over two lattice constants cubed will give depinning
thresholds in this range, suggesting that our choice for $j_d$ is possible.
The resistivity of the normal phase of NbCu has been estimated from
DC Residual Resistivity Ratio measurements of a Nb film on quartz.
\label{tab:parameters}}
\end{center}
\end{table*}

\begin{figure*}[!ht]
\begin{minipage}[t]{0.3\linewidth}
\centering (a) \par\smallskip
\includegraphics[width=\linewidth]{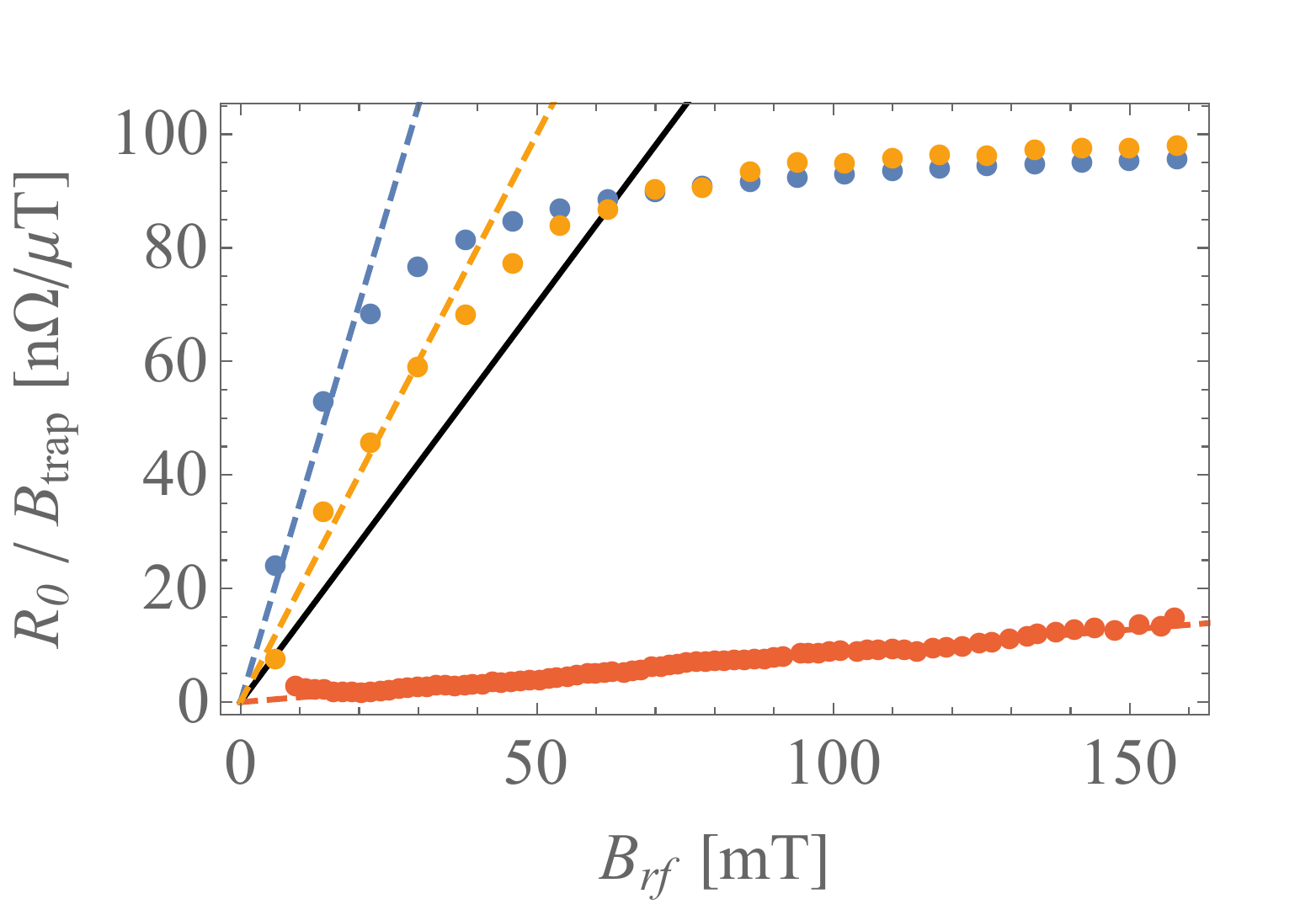}
\end{minipage}
\hspace{0.01cm}
\begin{minipage}[t]{0.3\linewidth}
\centering (b) \par\smallskip
\includegraphics[width=\linewidth]{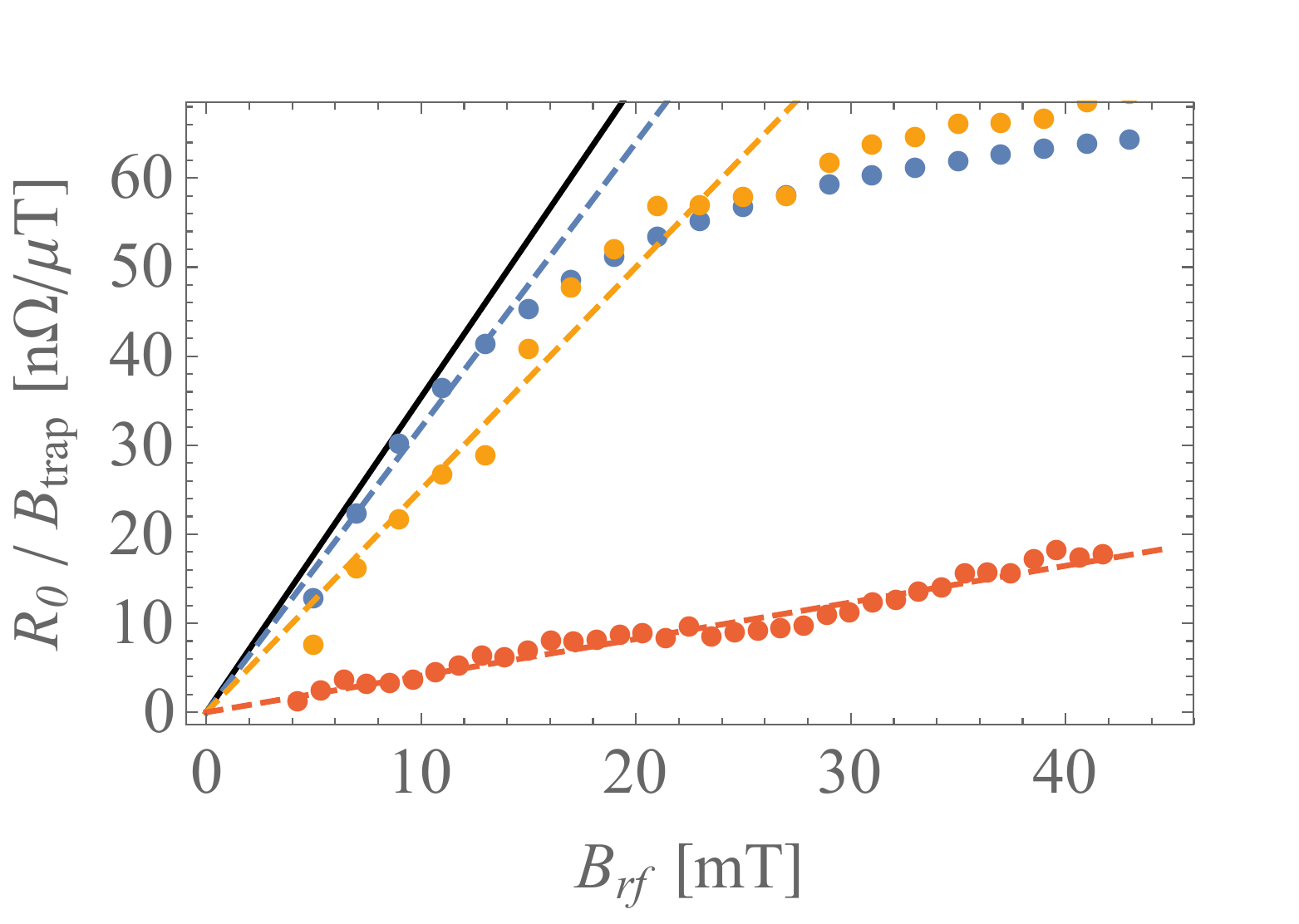}
\end{minipage}
\hspace{0.01cm}
\begin{minipage}[t]{0.3\linewidth}
\centering (c) \par\smallskip
\includegraphics[width=\linewidth]{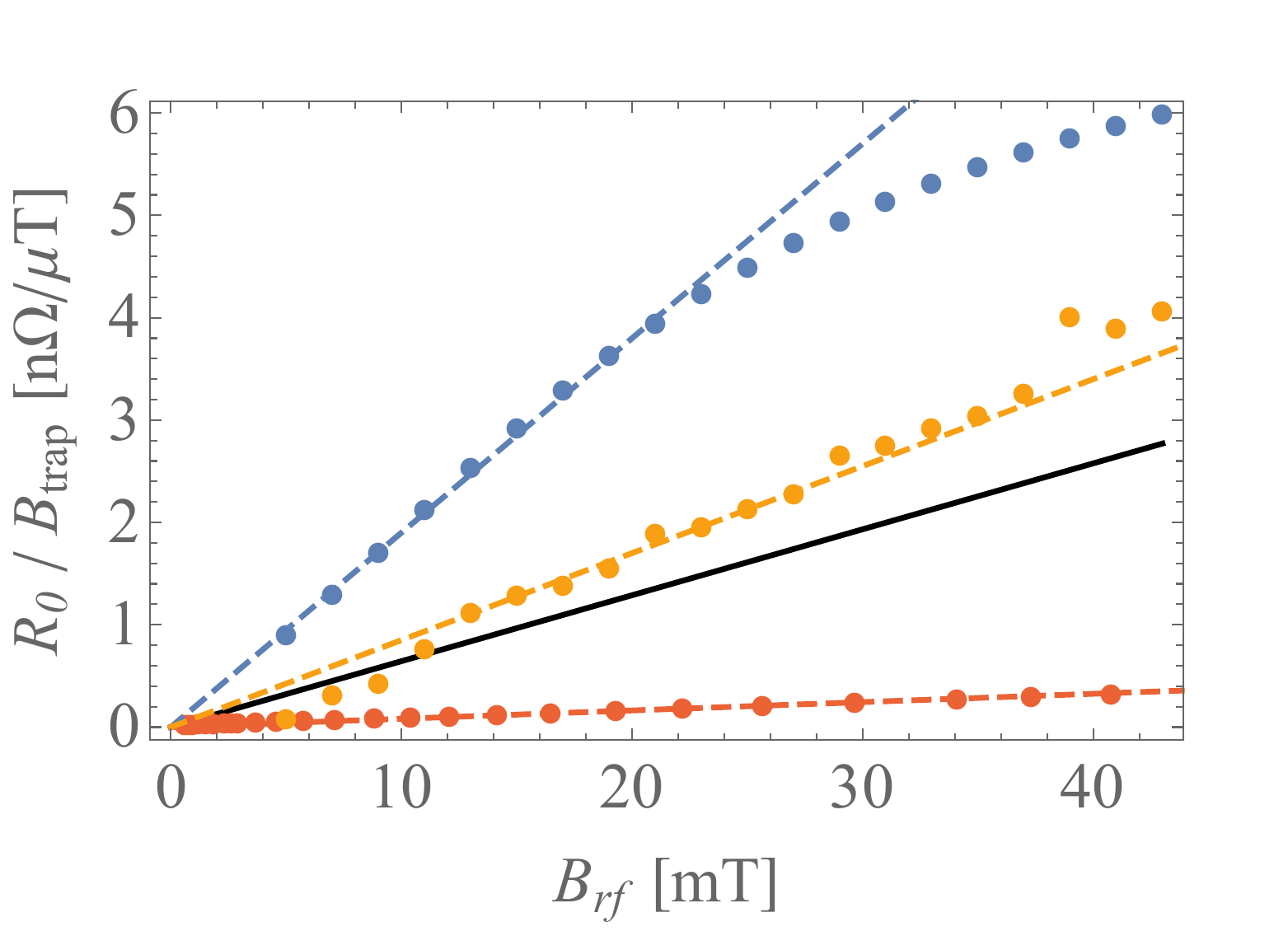}
\end{minipage}
\caption{Sensitivity of residual resistance to trapped flux as a function of the
rf field for doped Nb (a), Nb$_3$Sn (b) and NbCu (c) from experiments (red
circles), analytical calculations (black line), and numerical simulations (blue
and orange circles).
Theory curves use $j_d = 10^{10}$A/m$^2$.
Note that we could obtain numerical agreement with the experimental data
if we use larger depinning currents in our calculations.
The value we use is that measured for bulk depinning in
NbCu~\cite{miyazaki18,miyazaki_TTCFermi_17} in~(c), which admittedly has a very different
morphology from the Nb cavity.
This value is comparable to bulk pinning on dislocation cell structures in
Nb~\cite{santhanam76}; pinning on surface roughness (relevant here) could be
stronger especially in NbCu.
In the theory curves for Nb$_3$Sn, the value we use is the largest plausible
value from point-like impurities (Appendix~\ref{sec:checks}); pinning on
dislocations, grain boundaries, or tin-depleted regions would likely be 
stronger.
\label{fig:sensitivity}}
\end{figure*}

Figure~\ref{fig:sensitivity} shows a plot of the sensitivity of the
residual resistance to trapped magnetic flux as a function of the amplitude
of the rf field for doped Nb (a), Nb$_3$Sn (b) and NbCu (c).
Red circles correspond to experimental measurements, multiplied by
the correction factor $\mathcal{G}^{-1}\approx 2$ (for the Cornell doped-Nb
and Nb$_3$Sn cavities) and $\mathcal{G}^{-1}\approx 3$ (for the CERN
NbCu cavity~%
\footnote{%
Note that we have used the elliptical shape of the Cornell cavities in the
calculations of Appendix~\ref{sec:correction}. The correction factor for the
CERN NbCu cavity, which has a QWR geometry, might be different.
}), to account for vortex misalignment near the cavity equator
and field depletion near the cavity poles (see Appendix~\ref{sec:correction}).
We have also subtractted off the offset
($\lim_{B_\text{rf}\rightarrow 0} R_0/B_\text{trapped}$) of the measured
sensitivity, which is presumably due to loss mechanisms not involving
macroscopic vortex motion.
Blue and orange circles correspond to our numerical simulations using the
mean-field and the local-potential models, respectively (the dashed lines
emphasize the low-field linear behavior).
The black line corresponds to our approximated analytical solution given
by Eqs.~\eqref{eq:sensitivity_analytical} and~\eqref{eq:sensitivity_slope}.
Note that our calculations correctly capture the low-field linear behavior
observed in the experiments.
As expected, our calculations for the crossover field $B_X$, shown in
Table~\ref{tab:parameters}, are consistent with the simulation results for the
mean-field model (but consistently smaller than the crossover field of the
more realistic local-potential model.)
Also, note that we could fit the experimental data if we use larger
depinning currents (by a factor from about six for Nb$_3$Sn to twenty for
doped Nb) in our calculations.
The discrepancy between theory and experiments is larger for doped Nb
in part due to the low-frequency design (100MHz) of the NbCu cavity, and
the small coherence length of Nb$_3$Sn.
The remaining discrepancy could be ascribed to a number of factors,
which we discuss in Section~\ref{subsec:mechanisms}.
A word of caution: the fact that the analytical curve (black line) is close to
the local-potential solution (orange circles) in (a) and (c) should be taken
with a grain of salt.
The most \emph{realistic} model is the local-potential model.
The mean-field approach relies on a number of uncontrolled
approximations, and is particularly useful to provide order-of-magnitude
estimates and physical insights, rather than accurate predictions.

\subsection{Discrepancy between theory and experiment}
\label{subsec:mechanisms}

The theoretical curves in Figure~\ref{fig:sensitivity} use a depinning
current that is a factor of six to twenty too small to fit the experimental
curves.
The theory used the measured depinning current for one of the  materials
(NbCu), which by our estimates (Appendix~\ref{sec:checks}) is already too
high to be due to point-like pinning centers (impurity doping).
What could be the cause of the discrepancy?

As discussed in Section~\ref{sec:motion} and Appendix~\ref{sec:checks},
pinning on line-like impurities could be substantially stronger.
Indeed, vortex pinning on Nb dislocation cell structures is known to reach
values similar to those measured~\cite{santhanam76}.
Such pinning could be enhanced by impurity doping, if the dopant
preferentially segregated to the dislocation.
We would anticipate that the annealing steps in the preparation of the
niobium cavity would remove most of the dislocations. Pinning on grain
boundaries%
	\footnote{Pinning on tin-depleted regions in Nb$_3$Sn, or other 3D
		defects, would likely behave similarly.},
if it is not inescapable, would likely produce large depinning fields and a
residual resistance that depends on $B_\text{rf}$, but the grain size in
niobium is too large for our collective weak pinning theory to be applicable.
The role of dislocations or grain boundaries for Nb$_3$Sn is open for
further study.

But what about NbCu, where the pinning current was measured? 
Here the depinning current was deduced by measuring the hysteresis as the
external field was varied.
This adds a force per unit length to the whole vortex (a bulk measurement),
where the dissipation is due to a force on one end of the vortex.
Pinning due to surface roughness, or due to defects that arise more often
near the surface, could explain the discrepancy.
NbCu surfaces are particularly rough, as are the current Nb$_3$Sn surfaces.
Surface roughness, like grain boundaries, would likely not be modeled well
by collective weak pinning: each vortex would show little dissipation until
pushed hard enough to detach from its pinning site.
But  a distribution of vortex surface pinning strengths could generate a
field-dependent residual resistance.

Figure~\ref{fig:sensitivity} shows the theoretical and experimental residual
resistances per unit trapped flux.
Is it possible that the experimental value for the trapped flux is in error?
The  cavities were cooled very slowly in a DC applied field (to avoid forces
due to thermal gradients, which are usually maximized to expel flux%
~\cite{romanenko14,huang16,posen16}), and measurements show very
little flux expulsion from the cavity as a whole%
~\footnote{%
Note that near $T=0$, typical thermal gradients of $\sim T_c/$m result in
forces per length that are about $10^6$ smaller than the pinning force
used in our simulations for Nb.
Most of the flux expulsion must happen during cool-down, when $T$ is near
$T_c$, since the depinning force vanishes as
$(T_c-T)^{(5/12)(6-\mathcal{D})}$, according to GL theory.
Flux expulsion by thermal gradients is still a topic of general interest, and
deserves further investigation.
}.

However, recent measurements~\cite{posen18,martinello18} show large
heterogeneity in the heating due to trapped flux, both on the centimeter
scale of the detector resolution and on the decimeter scale of the cavity.
(The macroscale variations break the azimuthal symmetry, so are not due
to the geometrical factors discussed in Appendix~\ref{sec:correction}).
The simple theoretical picture of a uniform density of vortices independently
oscillating with a single pinning strength is clearly inapplicable.
The hypothesis that the flux remains homogeneous would demand that the
cold regions have much larger pinning strength than the hot regions, which
seems tentatively unlikely since the grain sizes are larger in the cold regions
(perhaps also indicating fewer dislocations within grains), and also the losses
increased when dislocations were added deliberately~\cite{posen18b}.
The fact that the flux is not expelled from the cavity as a whole does not
preclude the motion of flux within the cavity, either macroscopically or
microscopically.
If the vortices move, they either cluster into the hot regions, or they move
within the cold regions to nearby traps where they are strongly pinned.
The residual resistance due to the remaining vortices subject to collective
weak pinning would be linear in $B_\text{rf}$, with magnitude proportional by
$\langle j_d^{-1}\rangle$.
This motion to higher pinning would tend to reduce the dissipation per vortex.
Also, a substantial fraction of the vortices moving to sites where they are
inescapably and  rigidly pinned (and hence not dissipating) could explain the
discrepancy.

Measurements of the heterogeneity in the trapped flux would be useful.
Macroscopically, is there more trapped flux in the hot regions?
Microscopically, are the vortices trapped at grain boundaries or other
structures?
Is the pinning dependent on the grain orientation (and hence the orientation
of the screw dislocations, dominant in BCC metals)?
Is it dependent on the misorientation between grains?
Answering these questions could be of practical use.
Single-crystal cavities have been tried, but without controlling the surface
orientation.
One could also vary the grain orientation distribution or `texture' by suitable
plastic deformation before the final cavity is stamped into shape.
In doped Nb, the goal likely is to reduce all pinning and to maximize thermal
gradients during cooling to expel the flux.
In Nb$_3$Sn films grown on Nb and Nb films grown on Cu, thermal gradients
cause thermoelectric currents which induce trapped flux, so slow cooling is
necessary -- perhaps making stronger pinning beneficial.
This issue deserves further study.

\section{Final Remarks}
\label{sec:conclusions}

We have used the a model of vortex dynamics and collective weak
pinning theory to study vortex dissipation in superconductors.
We then applied our analysis to experiments performed in 1.3GHz
Nb$_3$Sn, doped Nb, and 100MHz NbCu cavities.
Using simple analytical calculations and standard numerical
simulations, we describe the low-field linear regime of the sensitivity
of the residual resistance to trapped magnetic flux.
Our results agree well with experiments performed in CERN and
Cornell.
We define a crossover field $B_X$, which increases with both
inverse frequency and depinning current, and that marks a transition
from a regime hysteretic to viscous-dominated losses.

We propose the tuning of material parameters as a method to
minimize the crossover field and reduce power dissipation in SRF
cavities.
Our simple approximated formulas for the slope of the sensitivity to
trapped flux (Eq.~\eqref{eq:sensitivity_slope}) and the crossover
field (Eq.~\eqref{eq:crossoverB}) provide a systematic way to
control and shed light into hysteretic-dominated trapped-flux
dissipation in SRF cavities.
The slope $A$ and the crossover field $B_X$ scale as
$f \, \lambda^2 / j_d$ and $(\rho_n / f)^{1/2} ( j_d / \kappa )$,
respectively.
As it should be anticipated, high-$f$, high-$\lambda$ and low-$j_d$
cavities yield large dissipation.
It would be interesting to apply our analysis to other Nb systems,
such as the Fermilab N-doped Nb cavities~\cite{checchin18},
and to adapt or extend our theory in view of the exciting (ongoing)
research developments on thermal flux expulsion, heterogeneous
flux trapping and the role of extended defects such as dislocations
and grain boundaries.

\appendix

\section{Sanity checks}
\label{sec:checks}

Here we discuss some approximations and sanity checks that are
associated with the derivation and analytical solution of the mean-field
model.

\begin{figure}[ht!]
\includegraphics[width=0.8\linewidth]{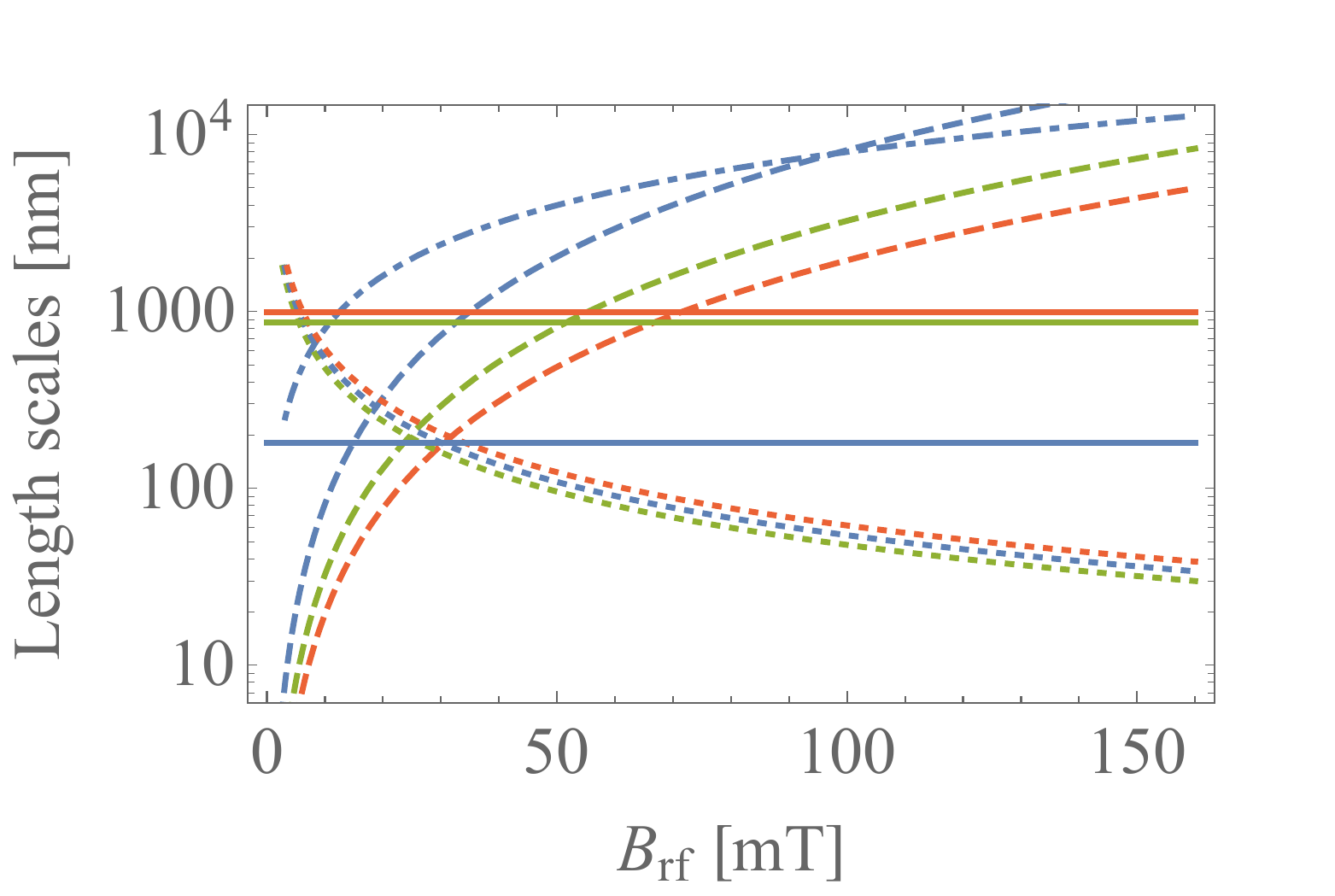}%
\caption{
Mean-field analytical calculations for the amplitude of deformation in
the $y$ (dashed) and $z$ (dash-dotted) directions, the radii of curvature
(dotted), and the depinning length (solid), for doped-Nb (green), Nb$_3$Sn
(blue) and NbCu (red) superconductors.
\label{fig:lengths}}
\end{figure}

We begin with a discussion of the characteristic length scales that
corroborate the collective weak pinning scenario and the point-force
approximation.
Let $a_y$ and $a_z$ be the vortex amplitudes of deformation in the $y$
and $z$ directions, respectively.
We can use the solution derived in Section~\ref{subsec:mf} to show
that,
\begin{equation}
	a_y
		\equiv \lambda \, \tilde{u} \, (\tilde{z}=0; \tilde{t}=1/4)
        	= \frac{\kappa}{\sqrt{2} \, c\,(\kappa)}
			\left(\frac{B_{rf}}{B_c}\right)^2 x_d,
	\label{eq:ay}
\end{equation}
where
\begin{equation}
	x_d \equiv \frac{B_c}{\mu_0 \, j_d},
	\label{eq:xd}
\end{equation}
is a characteristic length $\propto \lambda \, j_o /  j_d$.
The amplitude in the $z$ direction is given by
\begin{equation}
	a_z
       	\equiv \left. \lambda \, \tilde{z} \right|_{\tilde{u}(1/4)\rightarrow 0}
        	=  \frac{B_{rf}}{B_c} \, x_d.
    	\label{eq:az}
\end{equation}
Also, the curvature radius of the vortex line at $z=0$ is given by
\begin{equation}
	r_c
		= \frac{\lambda}{\tilde{u}^{\prime \prime}|_{\tilde{t}\rightarrow 1/4}}
		= \frac{c \, (\kappa)}{\sqrt{2} \, \kappa} \, x_d.
	\label{eq:rc}
\end{equation}
To restore the physical boundary condition at $z=0$ ($du/dz=0$), we
ad hoc bend the vortex line over a distance $\lambda$ from the
surface, so that $|u^{\prime \prime}| \approx |u^\prime| / \lambda$.
The curvature radius at $z=0$ then becomes
\begin{equation}
	r_\lambda
		= \frac{\lambda}{|\tilde{u}^\prime|_{\tilde{t}\rightarrow 1/4}}
		= \frac{c \, (\kappa)}{\sqrt{2}} \frac{B_c}{B_{rf}} \, \xi.
	\label{eq:rlambda}
\end{equation}
For completeness, Eqs.~\eqref{eq:LcSq} and~\eqref{eq:xd} result in
\begin{equation}
	{L_c}^2
		= \frac{\sqrt{2} \, c \, (\kappa)}{\kappa} \, x_d \, \xi.
	\label{eq:lc}
\end{equation}
Figure~\ref{fig:lengths} shows our mean-field solutions for $a_y$
(dashed curves), $a_z$ (dash-dotted), $r_\lambda$ (dotted), and $L_c$
(solid) for doped-Nb (green), Nb$_3$Sn (blue) and NbCu (red)
superconductors (note that all materials have the same $a_z$.)
For all three materials, the collective weak pinning assumption
$L_c \gg \xi$ is safely satisfied.
At large fields, the radii of curvature become small, and the
amplitudes of motion become large, thus justifying the point-force
approximation.
Note that the transverse amplitudes of motion ($a_y$) lie above the
micron scale for fields above $B_{rf} \approx 30$-$70$mT.
Grain sizes of Nb$_3$Sn are of order 1$\mu$m, emphasizing the role
played by extended defects in this case.

\begin{figure}[ht!]
\includegraphics[width=0.8\linewidth]{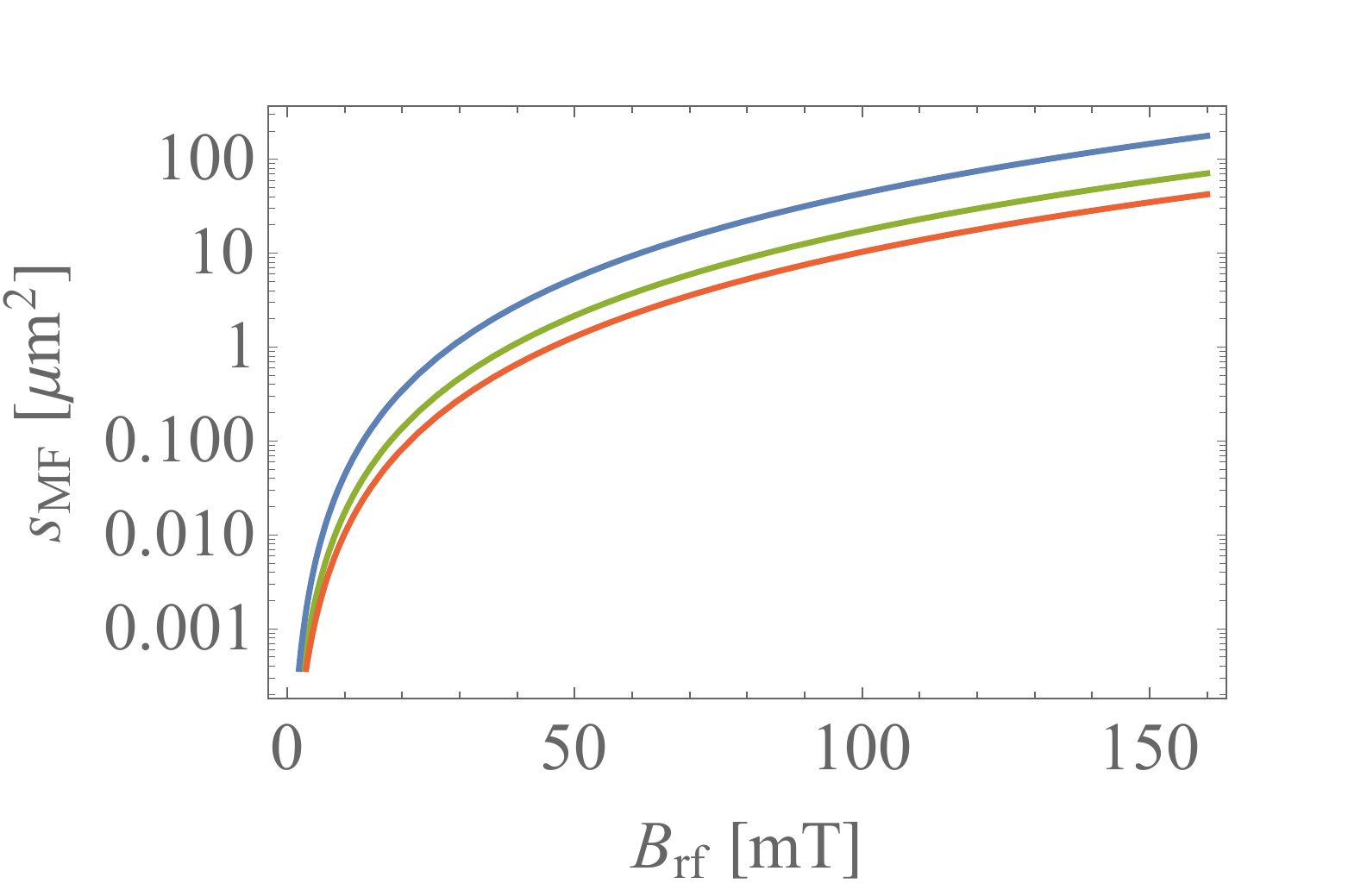}%
\caption{
Mean-field calculation of the area swept by one vortex oscillation
in the $y$-$z$ plane for doped Nb (green), Nb$_3$Sn (blue) and
NbCu (red) superconductors.
\label{fig:area}}
\end{figure}

Next we discuss the area swept by the vortex oscillations to justify
our assumption of independent vortex lines.
The area $s_\text{MF}$ in the $y$-$z$ plane that is swept by each
vortex oscillation is related to the average dissipated power per vortex
$ P_1/s_\text{MF} = 2 \, f \, f_P$, and is given by,
\begin{equation}
	s_\text{MF}
		= 2 \lambda^2 \int_0^{\infty} \tilde{u}_(\tilde{z}; 1/4) \, d\tilde{z}
			\nonumber \\
		= \frac{\sqrt{2}\, \kappa}{3\,c\, (\kappa)}
			\left(\frac{B_{rf}}{B_c}\right)^3 {x_d}^2.
\end{equation}
Figure~\ref{fig:area} shows a plot of $s_\text{MF}$ as a function of
$B_{rf}$ for doped-Nb (green), Nb$_3$Sn (blue) and NbCu (red)
superconductors.
Note that $s_\text{MF}$ approaches 1$\mu$m$^2$ at high fields,
which is about the grain size of typical Nb$_3$Sn, suggesting
that discrepancies with experiments might arise due to the vortex
interaction with grain boundaries.
On the other hand, from $B_\text{trap} / \phi_0 = N /s$, we
estimate a density of one vortex per $10^4$-$10^3 \mu$m$^2$
for a trapped magnetic induction of about $5$-$50$mG, suggesting
that the approximation of non-interacting vortices is consistent.

We end this section with a discussion about the relationship
between the depinning current and the density of impurities, and
the high depinning current used in our simulations.
Here we use Eqs.~\eqref{eq:LcSq} and~\eqref{eq:LcCube} to
eliminate $L_c$, and derive a formula relating the density of
impurities $n_\mathcal{D}$, the individual pinning force $F_i$, and
the depinning current $j_d$,
\begin{equation}
	{n_\mathcal{D}}^2
		= \frac{2 \, \epsilon_\ell \, {\phi_0}^3 {j_d}^3}{{F_i}^4\,
			\xi^{3-2\mathcal{D}}}.
	\label{eq:nD}
\end{equation}
We estimate the individual pinning force from the condensation
energy gained to move a vortex line from the border to the center
of a defect potential well of size $\xi$, i.e.
\begin{equation}
	F_i
		\approx a^{3-\mathcal{D}} \xi^{\mathcal{D}}
			\frac{{B_c}^2}{2 \, \mu_0 \,\xi},
	\label{eq:individualForce}
\end{equation}
where we have assumed that the impurity destroys
superconductivity over the volume
$a^{3-\mathcal{D}} \xi^{\mathcal{D}}$, with $a$ of order of an
atomic size.
Plugging~\eqref{eq:individualForce} back into~\eqref{eq:nD}
results in
\begin{equation}
	n_\mathcal{D}
		\approx 32 \times 2^{1/4} \pi^2 \sqrt{c\,(\kappa)}
			\left(\frac{a^2}{\xi}\right)^\mathcal{D}
			\left(\frac{\xi}{a^2} \sqrt{\frac{\lambda}{x_d}}\right)^3.
	\label{eq:pinning_estimate}
\end{equation}
We use Eq.~\eqref{eq:pinning_estimate} to estimate the density
of point-like impurities from the depinning current for a range of values
of the atomic distance $a$.
For Nb$_3$Sn, we find a density of 2--130 Nb atoms per impurity
for $a\sim1$-2 unit cell lengths, and a mean-free-path of $\sim1$nm,
where we have used BCS formulas for the dependence of
$\lambda$ and $\xi$ on mean free path~\cite{orlando79}.
Notice that this estimate is highly sensitive to the value of $a$, yet it
does not rule out the high depinning current that we have used if the
impurities affect a sufficiently large region.
On the other hand, our estimates suggest that high depinning currents
cannot be attributed to point-like impurities alone for doped-Nb and
NbCu.
Here we note that the term
$(a^2/\xi)^\mathcal{D}$ in~\eqref{eq:pinning_estimate} suggests that
consistent densities of defects can be associated with larger depinning
currents for \emph{extended} defects (with $\mathcal{D}>0$).
Additional experimental measurements of the depinning current
and mean free path might help test our assumptions using
collective weak pinning theory.

\section{Field-alignment correction}
\label{sec:correction}

\begin{figure}[ht!]
\includegraphics[width=0.8\linewidth]{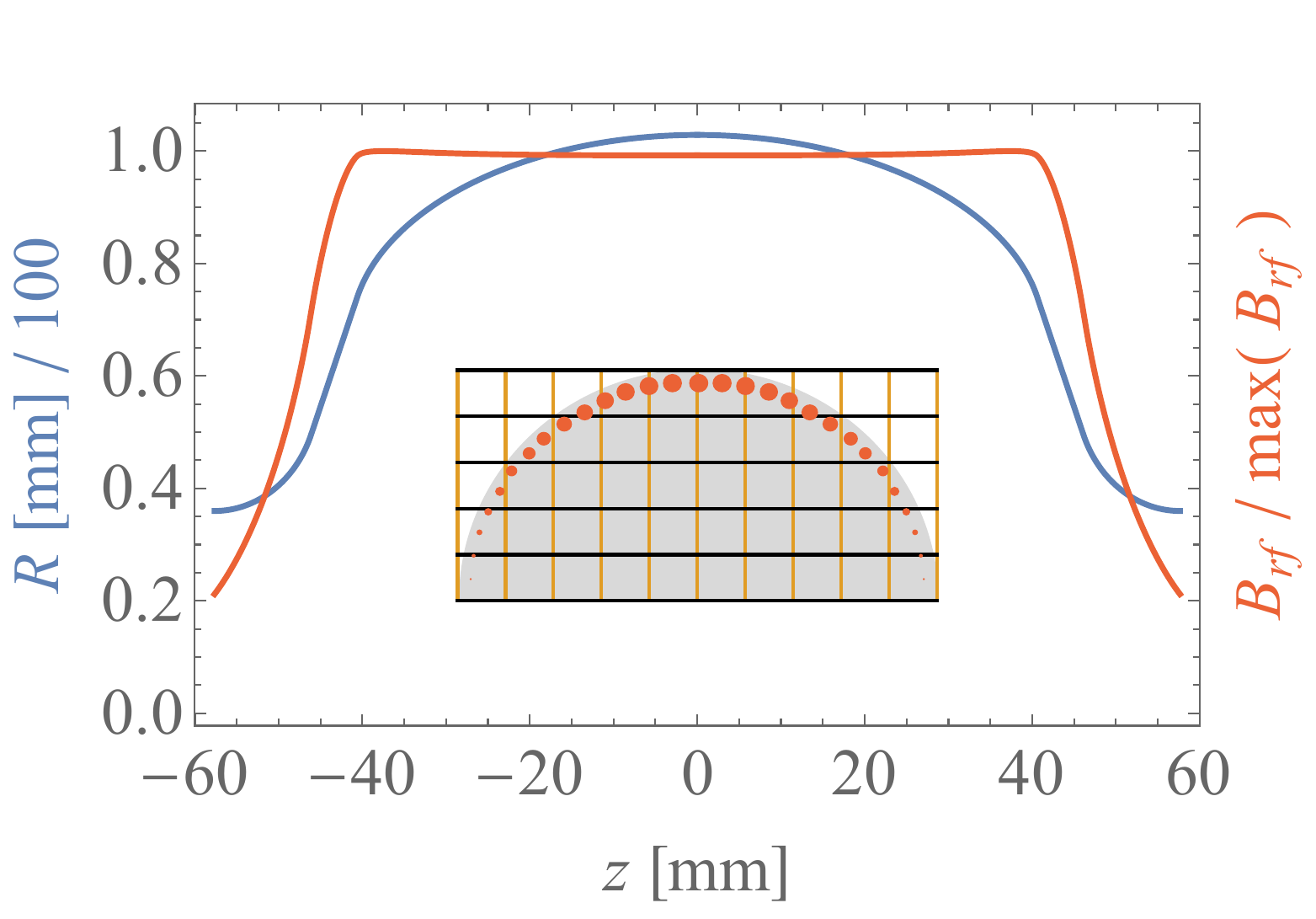}%
\caption{
Transverse radius (blue curve) and normalized amplitude of the rf
magnetic field (red) as a function of the longitudinal distance $z$ for
the Cornell 1.3GHz Nb$_3$Sn cavity.
The inset illustrates a similarly arranged cavity (gray disk), with red
circles corresponding to the rf field, and the black and yellow lines
corresponding to two possible directions for the DC field that creates
most of the trapped magnetic flux.
\label{fig:correction}}
\end{figure}

In our calculations of the residual resistance, we assume that
each vortex is initially perpendicular to the superconductor
surface, and is subject to the same value of the rf magnetic
field.
However, rf fields in real cavities are larger near the equator.
Figure~\ref{fig:correction} shows the normalized amplitude of
the rf magnetic field (red curve) and the cavity radius (blue) as
a function of the longitudinal coordinate $z$ (not to be mistaken
by the superconductor depth coordinate in the main text) for the
Cornell Nb$_3$Sn cavity.
The inset illustrates the upper portion of a similarly arranged
cavity (gray disk), with the red circles representing the rf
magnetic field at the surface (the field becomes smaller near
the poles), and the black and yellow lines representing two
possible directions for the DC magnetic field that creates
most of the trapped magnetic flux.
The black horizontal and the yellow vertical lines correspond
to the DC fields in the Cornell and Cern experimental setups,
respectively.
We then expect important corrections due to an interplay
between field depletion at the cavity poles and a non-uniform
density of vortices.

The density of vortices $\rho =\rho (z, \theta)$ for a DC magnetic
field $B_{DC}$ parallel or perpendicular to the $z$ axis is given by,
\begin{equation}
    \rho (z)
        \propto \frac{1}{\sqrt{1+{R^\prime}^2} } \times
            \begin{cases}{}
                |R^\prime |, & \text{for } B_{DC} \parallel z, 
                \nonumber \\
                |\cos \theta|, & \text{for } B_{DC} \perp z,
            \end{cases}
    \label{eq:dov}
\end{equation}
where $R^\prime \equiv dR / dz$, and $\theta$ is the polar angle
in cylindrical coordinates ($(R,\theta,z)$).
The surface area can be written as an integral over $z$ and $\theta$
of the ring infinitesimal area
$ds_\text{ring} = R\sqrt{1+{R^\prime}^2} \, dz \, d\theta$.
We also know the magnetic inductance $B_{rf}$ as a function of $z$.
In the region where the sensitivity to trapped flux increases linearly
with the rf field, the total dissipated power is proportional to
$\int {B_{rf}}^3 \rho \, R \, dz$.
In our model calculations, we have assumed $B_{rf}(z) = B_{rf} (0)$
and uniform $\rho$.
Thus, to make contact with the experimental results, we need correct
our predictions by a factor $\mathcal{G}$, defined as
\begin{equation}
    \mathcal{G}
        = \frac{\int {B_{rf} (z)}^3 \rho(z, \theta) R(z)
        	\sqrt{1+{R^\prime}^2} dz \, d\theta}{\int {B_{rf} (0)}^3 \cdot 1
		\cdot R(z) \sqrt{1+{R^\prime}^2} dz \, d\theta},
\end{equation}
where $\rho$ is given by Eq.~\eqref{eq:dov}.
Using the data shown in Fig~\ref{fig:correction}, we find
$\mathcal{G}=0.52$ and $0.37$ for  $B_{DC}$ parallel and perpendicular
to the $z$ axis, respectively.
This correction makes our theoretical prediction closer to the
experimental results.

\begin{acknowledgments}
We thank useful conversations with A. Gurevich, S. Posen, D. Hartill, and
S. Calatroni.
DBL and JPS were supported by the US National Science
Foundation under Award OIA-1549132, the Center for Bright
Beams.
DH, PNK and ML were supported by DOE Award No. DE-SC0008431,
and NSF Award No. NSF PHY-1416318 and  NSF PHY-1734189.
\end{acknowledgments}

\end{document}